\documentclass[a4paper,10pt]{article}
\usepackage[T1]{fontenc}
\usepackage[utf8]{inputenc}
\usepackage[italian, english]{babel}
\usepackage[autostyle,italian=guillemets]{csquotes}
\usepackage{multicol}
\usepackage{microtype}
\usepackage{guit}
\usepackage{booktabs}
\usepackage{graphicx}
\usepackage{mathptmx}
\usepackage{amsfonts}
\usepackage{amssymb}
\usepackage{amsmath}
\usepackage{ifpdf}
\usepackage{dcolumn}
\usepackage{textcomp}
\usepackage{tabularx,array}
\usepackage{ulem}
\usepackage{placeins}
\usepackage{epstopdf}
\usepackage{etex}
\date{}
\title{\textbf{Relationship between Climate and Famine from 1250 to 1350 in Central Italy}} 
\author{Tasselli, D. \\ TS Corporation - Astronomy and Astrophysics Department\\ Via Rugantino, 71 - 00169 Roma RM - Italy \\ E-mail: diego.tasselli@tscorporation.org \\ \\ Ricci, S. \\ TS Corporation - Meteorogical and Climatic Change Department\\ Via Giordano Bruno, 3 - 73057 Taviano LE - Italy \\ E-mail: stefano.ricci@tscorporation.org \\ \\ Bianchi, P. \\ TS Corporation - Geology and Geophysics Department\\ Via Reclusorio, 38 - 06138 Foligno PG - Italy \\ E-mail: pamela.bianchi@tscorporation.org \\ \\ Stanca, M. \\ TS Corporation - Astronomy and Astrophysics Department\\ Via Rugantino, 71 - 00169 Roma RM - Italy \\ E-mail: manuela.stanca@tscorporation.org \\ \\ Belli, R. \\ TS Corporation - Biological and Pharmaceutic Department\\ Via Cavalcanti, 10 - 51010 Massa e Cozzile PT - Italy \\ E-mail: rossella.belli@tscorporation.org}
\date{}
\begin{document}
\maketitle
\begin{abstract} 
This study aims to present an overview of how climate change brought about by events such as earthquakes, volcanic eruptions, flares and variations in solar activity, etc. have characterized the area of central Italy, analyzing and comparing astronomical, geological, meteorological, seismological, historical and climatic data, taking as reference some specific locations in the period between 1250 and 1350.
The ultimate goal is to understand how long-term climatic and geological events in this geographical area have changed the territory both from a natural and anthropic point of view, leading to highlight the occurrence of extreme events such as wave of plague of the period 1347-1351.
The analysis is performed using a statistical-historical approach and particular attention is paid to minimize any effect due to the error in the event of a lack of data. 
\end{abstract}
{\footnotesize This paper was prepared with the \LaTeX \\} 
\\ \\ 
{\normalsize}
\begin{multicols} %
	{2}
\section{Preface} 
Over the last 100,000 years, countless times the average temperature of the globe has fluctuated as a result of apocalyptic volcanic eruptions, solar flares, variations in the activity of the Sun, etc., which have led to local extreme events in different areas of the planet since meteorological point of view. From the most memorable (Toba volcano - Sumatra), which occurred only 74,000 years ago which reduced the human species to a few thousand individuals, up to the most recent (Pinatubo volcano - Philippines) which occurred in June 1991, which caused a lowering of the average temperature of the planet of 0.3$^{\circ}$C.\\ According to the most up-to-date research, over 1500 volcanoes have erupted in the last 10,000 years, but although many of these have a very low VEI classification (Volcanic Explosivity Index, measures the volcanic explosivity index, see figure 18 at the end of manuscript), a single violent eruption would be enough to affect the climate on a global scale.
\section{Introduction}
During the period of this study, between 1250 and 1350, natural events occurred worldwide which led to a mutation of the specific local balances of certain territories, both from a natural and anthropic point of view. Climate changes due to large volcanic eruptions have had a violent impact on the climate, even if lasting no more than a few years. In fact, volcanic ash has the ability to screen out sunlight about 30 times greater than the effectiveness of preventing the heat of the globe from escaping to the outside.
During the 3 years that the volcanic ash takes to settle on the ground after a violent eruption, the temperature of a part of the globe or of the whole globe can sometimes drop by a whole degree (it seems little but it is a lot). This leads, as a not negligible consequence, to the change of the climate at the local level, with extreme events of rains and/or storms, or from the opposite point of view, extreme droughts accompanied in some periods by epidemics. 
\section{Methodology of data acquisition and analysis}
All the scientific data have been obtained from searches in specific databases, while the historical data have been obtained with historical-literary analysis coming from the literature present in the bibliography. \\Finally, for each single region, the main cities were identified where the major events reported in this study occurred or were perceived.
\section{Link "Volcanic activity - Climate"}
But before analyzing this more accurately, it is necessary to briefly expose a physical process that underlies the climatic activity of our planet. This process is called \textbf{radiative equilibrium} or more commonly known as the \textbf{"sun-earth energy balance"}.\\
As the name suggests, this physical process describes how our planet exchanges energy in the form of radiation with outer space. In the radiative equilibrium it is possible to distinguish an incoming radiation flux (which is given by solar radiation) and an outgoing radiation flux (which is given by the thermal radiation that the terrestrial climate system emits towards space and which includes the infrared radiation emitted by the land surface that manages to pass through the atmosphere). 
Of the incoming flow, or the solar radiation incident on our planet, about 70\% is absorbed (by the earth's surface and atmosphere), while the remaining 30\% is reflected back by the planet. It therefore follows that there is no real equilibrium between incoming and outgoing flux, but the outgoing flux will be less than the incoming flux, and therefore in quantitative terms of the 342 W/m$^{2}$ of solar radiation incident on the planet, only about 1/3 (i.e. 112 W/m$^{2}$) will be reflected back to space.
\\ In order for there to be such a radiative equilibrium to allow life as we know it, the outgoing infrared radiation must have a global average value of about 112 W/m$^{2}$. \\ \\
\includegraphics[width=1\linewidth]{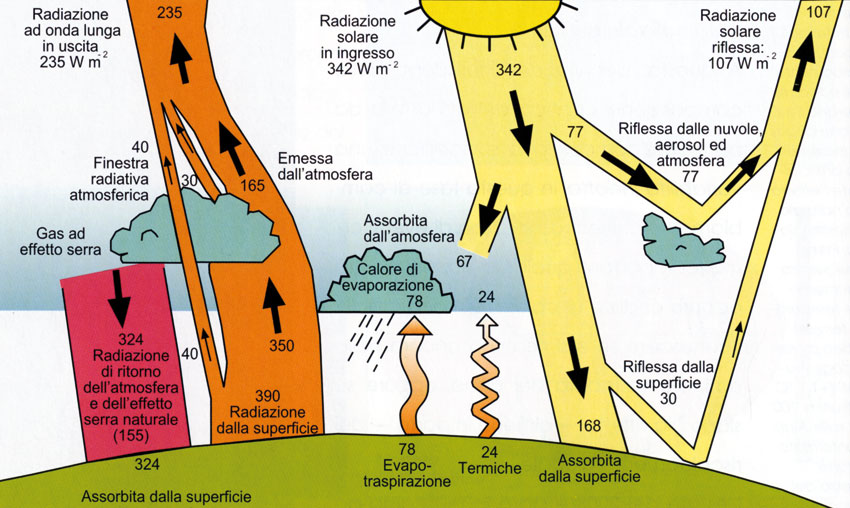}
{\scriptsize \textit{Fig.1: Schematic diagram of the energy balance between the Sun, Earth and atmosphere}}. $^{\scriptsize \cite{RVA:ref1}}$ \\ \\
Any process that alters the equilibrium values between the incoming and outgoing radiation flux, would cause a radiative forcing on the system which can be positive or negative.
To understand the importance of the radiative equilibrium, just think that if this physical process were absent, the average global temperature of our planet would not be around the "today's"15$^{\circ}$C, but at -20$^{\circ}$C.\\ \\
\textbf{Positive forcing:}
\begin{itemize}
 \item If the incoming flow increases;
 \item If the outgoing flow decreases (greenhouse effect).
\end{itemize}
\textbf{Result}: increase in average atmospheric temperature\\ \\
 \textbf{Negative forcing:}
\begin{itemize}
 \item If the incoming flow decreases;
 \item If the outgoing flow increases.
\end{itemize}
\textbf{Result}: decrease in average atmospheric temperature \\ \\ 
Therefore, in order for a negative forcing to occur such as to cause a decrease in the global average temperature, a series of processes should occur that would decrease the incoming flow (therefore absorbed by our planet), or increase the outgoing flow. \\ Technically speaking, the way more plausible would be that of the decrease in solar radiation absorbed by the planet (decrease in the incoming flux). This would be possible only if the planet reflected more solar radiation, thus recording an increase in the planetary albedo: in part of the surface albedo (possible if there was an increase in snow), but above all an increase in the albedo of clouds (possible if there was an increase in cloud cover consisting mainly of clouds rich in sulphates, and an increase in the optical thickness of the cloudy body).
\\ From this we understand how particles made up of sulphates (of marine, wind, anthropic origin), with dielectric properties such as to produce strong scattering (diffusion of the electromagnetic wave) and weak absorption of solar radiation, would increase the albedo of the clouds (if present in them), as opposed to carbonaceous particles (which if present would decrease the albedo of the cloudy body). \\
As happened in the past, huge eruptions of volcanoes would create volcanic clouds such as to modify the chemical composition of the clouds, increasing the quantity of sulphates, and consequently increasing the planetary albedo and causing a negative forcing of the radiative equilibrium. Being a volcanic cloud consisting not only of sulphates but also and above all of: 
\begin{itemize}
 \item	H$_{2}$o (on average 60\%) - Water;
 \item	Co$_{2}$ - Carbonic Anhydride;
 \item	Co - Carbon Monoxide;
 \item	Hcl - Hydrochloric Acid. 
\end{itemize}
in case of volcanic eruptions so violent as to create a column of volcanic cloud so high as to reach the lower stratosphere, the stratospheric content of Aerosol particles could vary, strongly affecting the terrestrial climate. After a couple of months the eruption (or rather the volcanic cloud) would extend over a wide latitudinal belt, and in the aforementioned cloud there would be a concentration of droplets composed of 75\% sulfuric acid (H$_{2}$SO$_{4}$) and 25\% by H$_{2}$O. \\
These particles, depending on the intensity of the volcanic eruption, would remain for months or years in the stratosphere before falling, thus increasing the albedo of the clouds. \\
Furthermore, with an increase in volcanic clouds, and therefore an increase in the sulfuric acid with which the amines present in the atmosphere can combine $^{\scriptsize \cite{CERN:ref1}}$, the formation of condensation nuclei around to which the water vapor would thicken, favoring the formation of new clouds that will reflect the incident solar radiation thus causing a further increase in planetary albedo. \\
This makes us understand how the particles present in a volcanic cloud, in particular the sulphates, increasing the albedo of the clouds would cause a general global cooling. \\ But the sulphates would affect the global climate not only by increasing the albedo of the cloud bodies. , but indirectly by destroying the ozone layer. In fact, they are able to easily convert chlorofluorocarbons (CFCs) into much more active compounds and capable of accelerating the destruction of the ozone layer. Consequently, a loss of ozone would cause stratospheric cooling, which in turn would entail a strengthening of the polar vortex in the stratosphere and (most likely) in the troposphere, such as to initially increase the extent of the glaciers (and therefore increase the superficial albedo and further the planetary albedo) if the aforementioned baric figure were centered in its geographical pole, and later due to resistance in the troposphere that would make the polar vortex assume an axial shape, there would be a cooling at well latitudes lower, causing a further increase in superficial and therefore planetary albedo. \\
\includegraphics[width=1\linewidth]{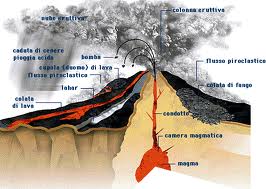}
{\small \textit{Fig.2: Diagram of a volcanic eruption}} \\ \\ Moved by the wind, the clouds move over large areas and sometimes reach the entire planet, having as a global effect a thick cloud cover, which the energy of the sun cannot cross. Furthermore, once volcanic clouds have fallen back (and therefore no longer suspended in the atmosphere) they would continue, albeit indirectly, to negatively force the radiative balance, causing an increase in cloud cover and consequently an increase in planetary albedo. To understand this latter process it is necessary to understand what atmospheric aerosol is. 
\subsection{L'atmospheric aerosol} it is composed of particles and corpuscles suspended in the atmosphere, the chemical nature of which is variable and depends on their origin. It can be released into the atmosphere by the action of the winds on the deserts, or by the action of the wind on the oceans (or by the volcanoes, as seen previously). 
 \begin{itemize}
 \item	aerosols constitute nuclei of aggregation for water vapor molecules, contributing to the formation of clouds;
 \item	aerosols play an extremely important role in the scattering process of solar radiation.
 \end{itemize}
The number of aerosol particles per unit volume decreases as the particle size increases. In particular, the particles of marine and wind origin have dielectric properties such as to produce strong scattering and weak absorption of solar radiation. The sea is among the first sources of global aerosol. \\ \\
\includegraphics[width=1\linewidth]{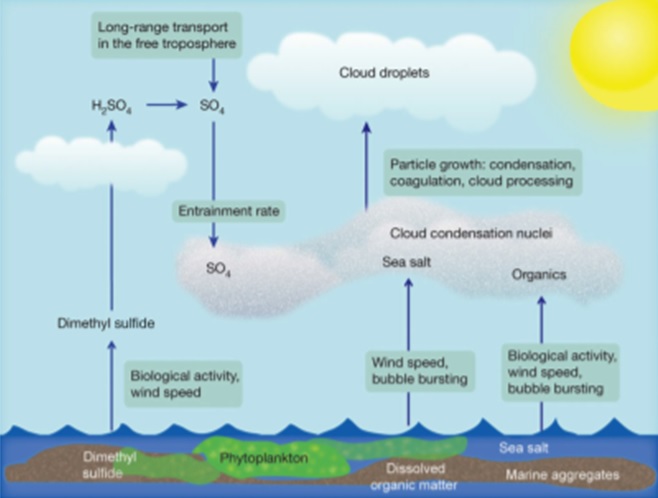}
{\scriptsize \textit{Fig.3: Schematic description of marine aerosol formation and processes (Quinn e Bates, 2011)}}.$^{\scriptsize \cite{aereosolo:ref1}}$ \\ \\
Marine aerosol (also called marine spray) originates from the action of winds on the surface of the water. The chemical composition ("classic") of marine aerosol is divided into fine fraction (diameter 1um), where over 90 \% of the fine fraction is made up of sulphate, while 100 \% of the coarse fraction is sea salt. \ \
Studies show that the physical and chemical properties of marine aerosol vary seasonally similarly to variations in marine biological activity, particularly the flowering of marine phytoplankton. The organic component (mainly consisting of insoluble organic compounds) dominates the fine fraction of the marine aerosol during the flowering period of marine phytoplankton. These organic particles are released into the atmosphere by the explosion of air bubbles produced by the wave motion of the oceans. \\
This new source of organic surfactant material increases the availability of condensation nuclei in the clouds of the atmosphere and therefore influences the planet's climate. \\
\textbf{In the period of low biological activity} in the coarse fraction sea salt dominates (about 100\%), while in the fine fraction, sea salt is present for about 75\% while only about 20\% is made up of organic material (soluble and insoluble).\\
\textbf{In the period of high biological activity}in the coarse fraction the sea salt dominates, even if there is an amount of organic material (soluble and insoluble) equal to about 4\%, while in the fine fraction the presence of organic material dominates (about 63\%) and in particular the quantity of insoluble organic material prevails over the quantity of soluble organic material. \\
Climate models currently consider marine aerosol to be composed solely of sulphate and sea salt. This new source of surfactant organic material, greater during the phytoplankton flowering period, increases the availability of cloud condensation nuclei in the atmosphere, thus also allowing for an increase in cloud cover. The flowering of phytoplankton occurs thanks to the sun (through which photosynthesis is possible) and the quantity of nutrients present in the water.\\
According to some studies, the concentration of phytoplankton would increase considerably in the presence of waters rich in iron. Thus, in this process the volcanic ash would act as a fertilizer for the oceans, releasing various substances (including iron) considered nutrients for phytoplankton. \\ It is also believed that phytoplankton can remove roughly the same amount of phytoplankton from the atmosphere. carbon dioxide absorbed by terrestrial vegetation, thus exerting a contrasting action against the greenhouse effect and global warming.
It is possible to assert that volcanic ash, acting as fertilizer for the oceans of our planet, would favor an increase in the concentration of phytoplankton, which in turn, through the dynamics described above, would (indirectly) cause an increase in cloud cover, a consequent increase in planetary albedo and finally a negative forcing in the radiative equilibrium.
\section{Geological data} 
This paragraph presents a geological-climatic analysis of the area falling within central Italy, divided by the seismoligical part, into the 4 major regions that were most affected by the events dealt with in this paper. \\The reference area is between the Appennince chain, formed between the upper Miocene and the lower Pliocene (when it assumed tts current form regressive from an open sea to a continental environment) and the Tyrrhenian Sea.\\ The main geological areas that have the greatest impact from the volcanological point of view for the area of this study are those characterized by the Districts: Vulsino, Cimino-Vincano, Tolfetano, Sabatino and Albano. \\ \\
\includegraphics[width=1\linewidth]{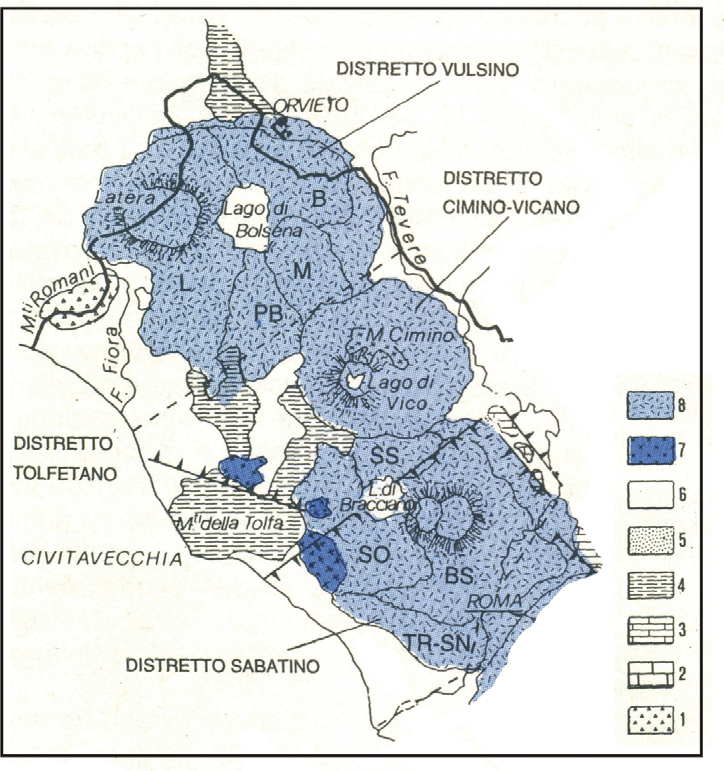}
\\ {\small \textit{Fig. 4: Map of volcanic complexes in the study area. Legend: 1: rocks of the metamorphosed base; 2: sediments from the Lazio-Abruzzo platform; 3: sediments of the Umbrian Marche basin; 4: allochthonous sediments of the Ligurian and subligur complex; 5: flyschoid allochthonous sediments; 6: neo-autochthonous sandy-clayey-gravelly sediments; 7: volcanic districts with acid to intermediate chemism; 8: volcanic districts with a potassic to highly potassic character (PB: Paleobolsena volcanic complex; B: Bolsena volcanic complex; M: Montefiascone volcanic complex; L: Latera volcanic complex; SW: western sector activity; SS: activity of the northern sector; BS: Sacrofano-Baccano complex; TR-SN: pyroclastic flow of red tuff with black slag) - (modified from Italian Geological Society,1993)}} $^{\scriptsize \cite{DeRita:ref1}}$ 
\subsubsection{Vulsini (Volsino) Volcanic District} 
The volcanic complex of the Vulsini (or Volsini), located in northern Lazio, extends for about 2.200 km$^{2}$, is a caldera with elevation of 800 mt. \\The main area formed about 0.3 and 0.16 million years ago, during the explosive eruptions of the Pleistocene (five large plinian fall deposits were erupted from the cones of the Caldera Latera or nearby during the Upper Pleistocene), which they contributed to generate the main complex, formed by the Bolsena caldera 16 km wide (filled by the waters of the lake of the same name), and the Latera caldera 8X11 km wide to the west.\\
The latest major eruption formed unwelded pumice flows and welded airfall tuffs of the Pitigliano Formation, associated with the collapse of the Vepe caldera some 166,000 years ago at the northwest end of the Latera caldera. \\Post-caldera volcanism produced scoria cones and lava flows from vents within and to the west of Latera caldera.\\
The area affected by the Vulsino volcanic district has been described and analyzed in detail in the work referring to the study of the Municipality of Montefiascone (VT) $^{\scriptsize \cite{Tasselli:ref2}}$. \\The following figure shows the structure of this volcanic complex. \\ \\
\includegraphics[width=1\linewidth]{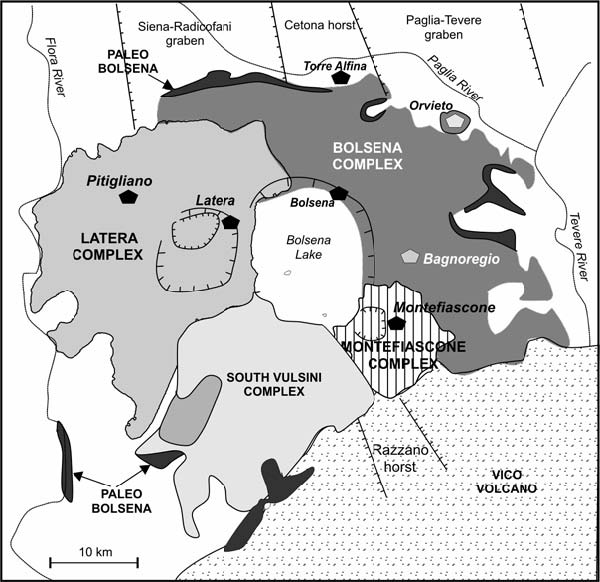} 
\\ {\small \textit{Fig.5: Schematic geological map of the Vulsini area. Image from:\\ https://www.alexstrekeisen.it/provincie/vulsini.php}}
\subsection{Colli Albani (Alban Hills) Disctrict}
The Colli Albani (Alban Hills or Laziale Volcano) is a large volcano layer with a central caldera, with elevation of 949 mt., located about 20 km south-east of Rome, developed in an area characterized by oriented extensional fault systems NW-SE, NE-SW e NS.\\ Volcanic products are essentially formed by pyroclastic deposits (peperino) minor lava flows, erupted in a time span between 0.6 Ma up to the Roman era. \\The volcanic rocks extend over an area of approximately 1000 Km$^{2}$ and cover Plio-Pleistocene marine deposits that fill the extensional basins, developed in flysh deposits and on carbonate platforms (facies of Sabina and Facies Umbro-Laziale).\\ The Alban Hills are located on the west side of the Ancona-Anzio tectonic line and the numerous volcanological, petrological and morphological studies have made it possible to reconstruct the main eruptive phases:
\begin{itemize}
\item \textbf{Tuscolano-Artemisio phase (from about 0.6 to 0.3 Ma):} In this phase the construction of the main volcanic building took place, which ended with a Calderico collapse;
\item \textbf{Phase of Faete (or phase of Campi Hannibal, from about 0.3 to 0.2 Ma):} In this phase there was the formation of the intracaldertic cone of Feate and subsequently an intracaldertic collapse. Faete's activity ended with the collapse of the nested cladera of Campi di Annibale and Strombolian circum-calderic eruptions.
\item \textbf{Hydromagmatic phase (from about 0.2 to 0.02 Ma):} This phase was characterized by violent phreatomagmatic eruptions. It developed along numerous volcanic centers located in the west of the main volcanic building. \\The best known eruptive centers are the maars of Nemi and Lake Albano. \\The younger ages, for the volcanic deposits of the Alban Hills, measured in the area of Lake Albano indicate an age below 19 Ka.
\end{itemize} 
The volcanic rocks of the Alban Hills have a strongly ultra-potassic composition, and strongly undersaturated; the products range from tephra to foidites (leucitites) and tephraphonolites.\\ \\
\includegraphics[width=1\linewidth]{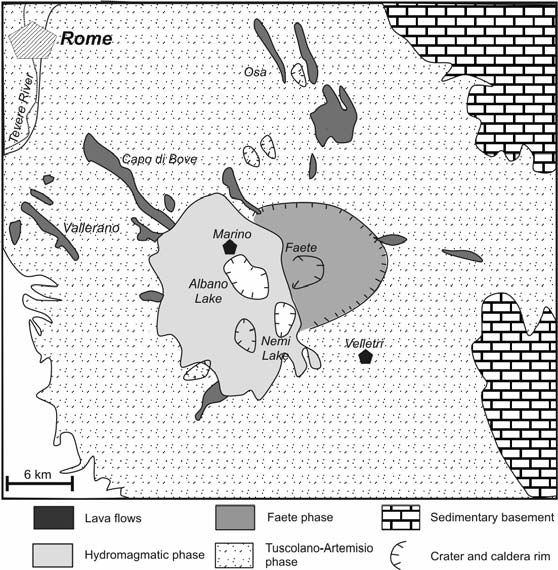} \\
{\small \textit{Fig.6: Geological structure of the complex of the Colli Albani (Alban Hills)}} $^{\scriptsize \cite{Tasselli:ref3} \cite{DeRita:ref1}}$ \\ \\
The lava flows are more mafic than the pyroclastic deposits coming from the same eruptive centers, but they maintain similar alkalinity values.\\
The pyroclastic deposits of the Alban Hills are made up of slag and pumice, with a porphyritic structure containing leucite and clinopyroxene crystals, and abundant xenoliths of various kinds are often found. \\
The current situation of the Alban Hills has highlighted volcanic activity even in Roman times (some Roman authors report strange phenomena such as stone rains, sudden fires, explosions), and there are numerous archaeological data (pottery and other pre-Roman human artifacts) found underneath pyroclastic deposits. \\These data indicate that the volcanic activity is considerably younger than the age obtained from the Albano lake (19 Ka). \\ Currently based on seismic and historical data, it is believed that the Alban Hills are in a phase of long quiescence.{\scriptsize $^{\cite{Peccerillo A:2005}}$} 
\subsection{Seismicity in Central Italy}
The regions affected by this study highlight the presence of many faults present mainly in the Umbrian-Marche-Abruzzese territory, as clearly shown in the following map: \\ \includegraphics[width=1\linewidth]{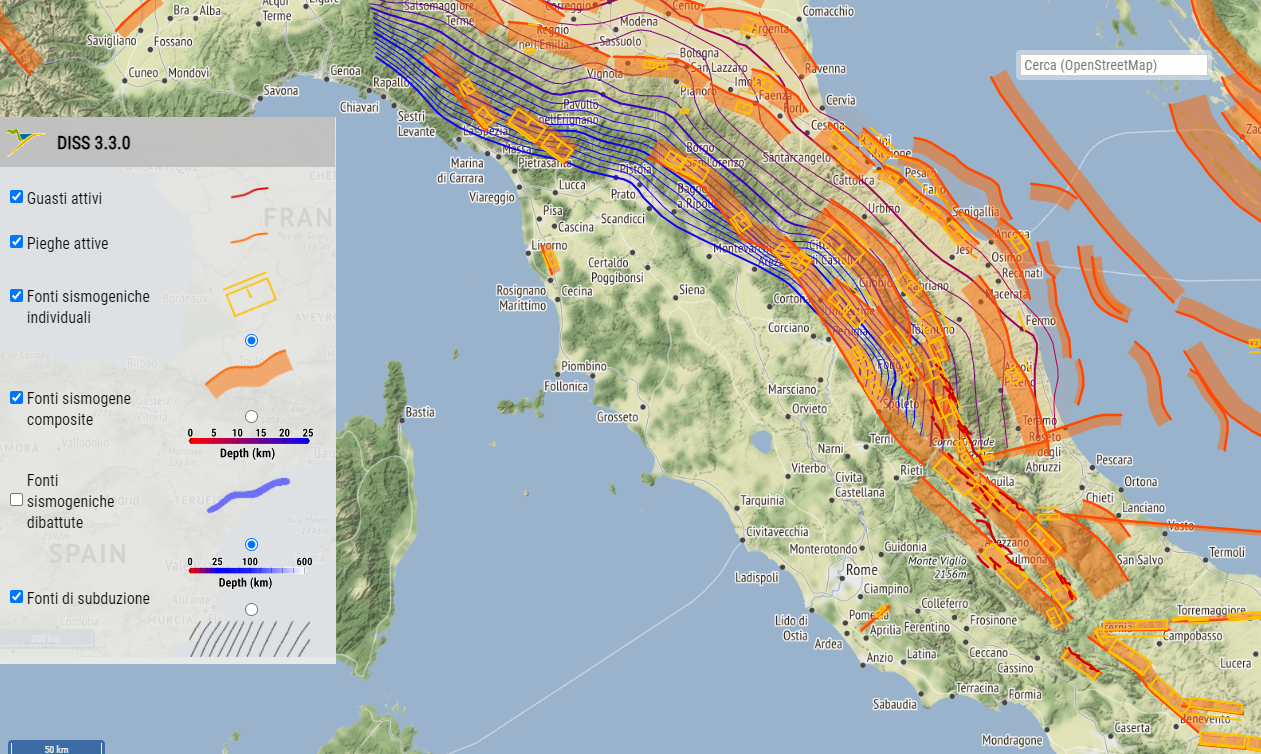}
{\small \textit{Fig.7: Map of the active faults in central Italy}} {\scriptsize $^{\cite{DISS:ref1}}$}
\\ Particularly affected by seismicity is obviously the Apennine belt which affects the eastern part of the Lazio region, the entire region of Umbria, and the western part of the Marche and Abruzzo regions.\\ Precisely in this particular area there were the main seismic events recorded in the period covered by this study, which in the following articulations, highlights the geological characterization of the individual regions involved in this study, recalling and referring to the specific studies indicated in the bibliography at the end of the item.  
\subsubsection{Lazio region} 
The Lazio region is characterized by different areas of seismicity. \\The Apennine area obviously has a classification of greater seismicity, and it is precisely from these areas and also from the Colli Albani area, from the area of the Vulsini, Cimini-Vicano, Tolfetani and Sabatini mountains that the major seismic events were generated that directly or indirectly, they affected the city of Rome (chosen as the reference city for this study).\\ The following figure highlights the new regional seismicity map:\\
\includegraphics[width=0.55\textwidth{}]{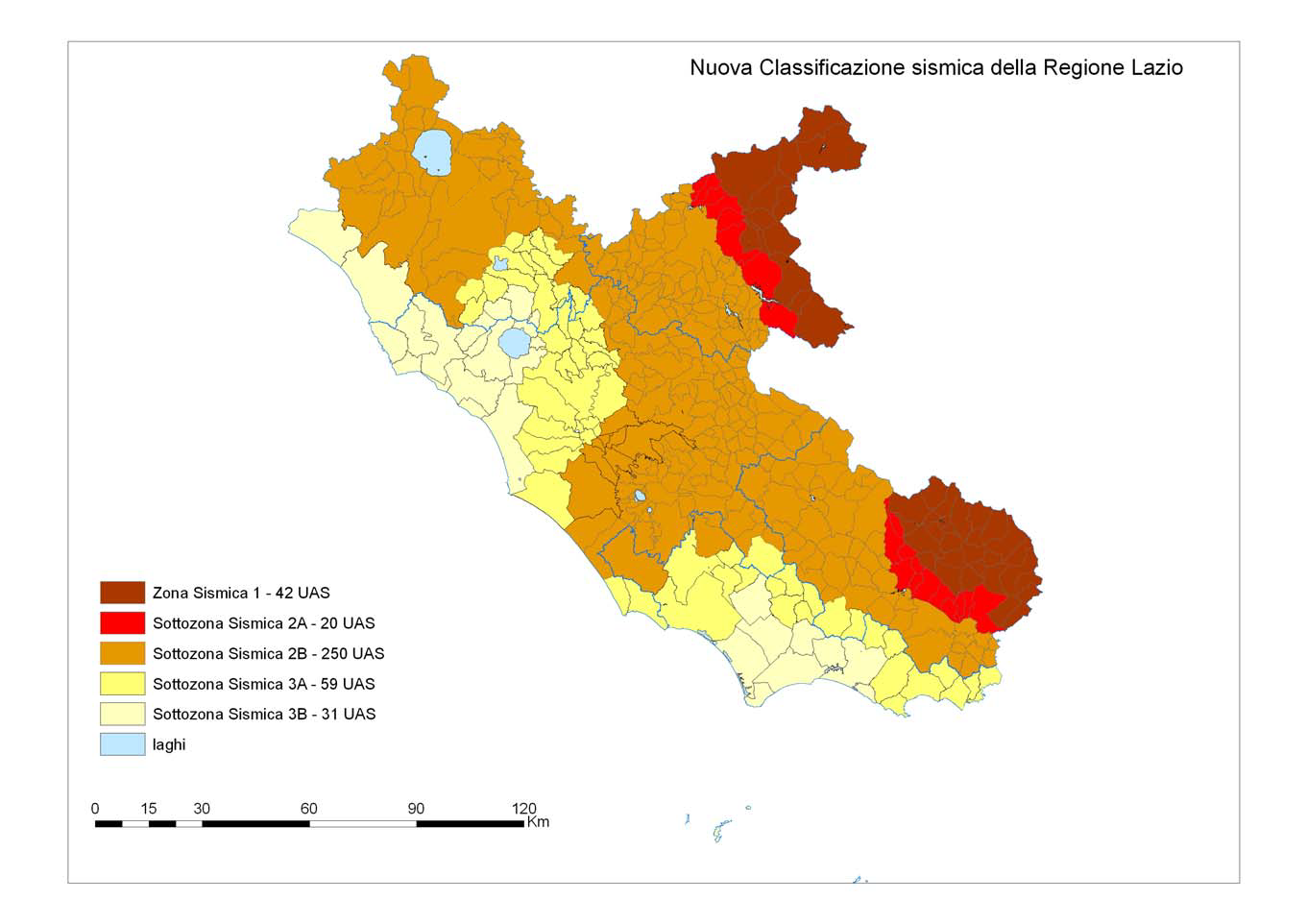} \\
{\small \textit{Fig.8: Lazio Seismic Hazard Map (MPS). The acceleration values refer to a return of 475 years (INGV2004). Legend:
Seismic zone 1 - 42 UAS - Seismic sub-area 2A - 20 UAS - Seismic sub-area 2B - 250 UAS -
Seismic sub-area 3A - 59 UAS - Seismic sub-area 3B - 31 UAS - Lakes}} {\scriptsize $^{\cite{INGV:2004}}$} \\ \\
The area examined by this study, identified by Sheet \textbf{"374 Rome"} includes the Tiber Valley, the area of the Albani volcanoes to the south and the Sabatini Mountains to the north, but for the most part it is located in the urban area of Rome.\\
The municipality of Rome, is characterized by an area with medium-high seismicity.\\ In fact in table A attached to the Decree of 14/01/2008 of the Ministry of Infrastructures the estimates of seismic hazard are highlighted and from these the elastic response spectrum (horizontal and vertical) of the seismic actions highlighted in the following graph has been determined: \\ \\
\includegraphics[width=0.49\textwidth{}]{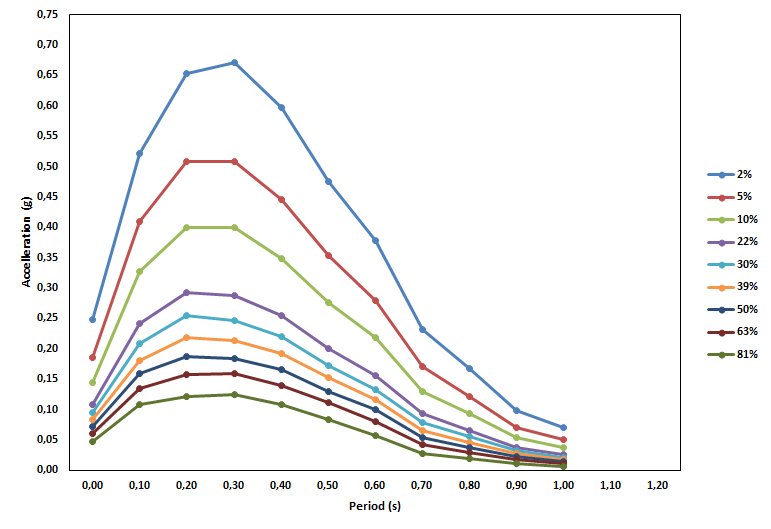} \\
{\scriptsize \textit{Fig.9: The spectra indicate the calculated shaking values for 11 spectral periods, ranging from 0 to 2 seconds. The PGA corresponds to the period of 0 seconds. The graph refers to the median estimates (50$^{\circ}$ percentile) proposed by the hazard model. The different spectra in the graph refer to different overflow probabilities (PoE) over 50 years.}} {\scriptsize $^{\cite{INGV:2004:ref2} \cite{Stucchi:ref1}}$} 
\subsubsection{Abruzzo Region}
The Abruzzo region is characterized by medium-high seismicity, above all in the Apennine area. \\The area examined by this study, identified by Sheet \textbf{"359 Aquila"}{\scriptsize $^{\cite{CNR:ref1}}$}, includes the central-northern sector of the Abruzzo Apennines, immediately south of the Gran Sasso. \\ The area is part of the Adriatic block with African affinities, in particular of the outermost segment of the central Apennines, of which it reflects both the geological characteristics and the palaeographic and tectonic evolutionary trend.\\
The sector was in fact structured by the tectonic superposition of units belonging to different paleogeographic domains, formed by discordant basins on the deformed substrate and sometimes incorporated in the chain domain..\\
The area was involved in the structuring of the Apennine chain starting from the lower Messian up to the lower Pliocene and sees the presence of active fault systems that give the area an organized structure in ridges and depressions bordered by faults.\\ 
The whole area is affected by intense neotectonic activity well highlighted by the high regional seismicity;\\ in fact Aquila and the surrounding areas have been affected several times by destructive earthquakes (the one of 1349 is highlighted in this study).\\ \\
\includegraphics[width=0.49\textwidth{}]{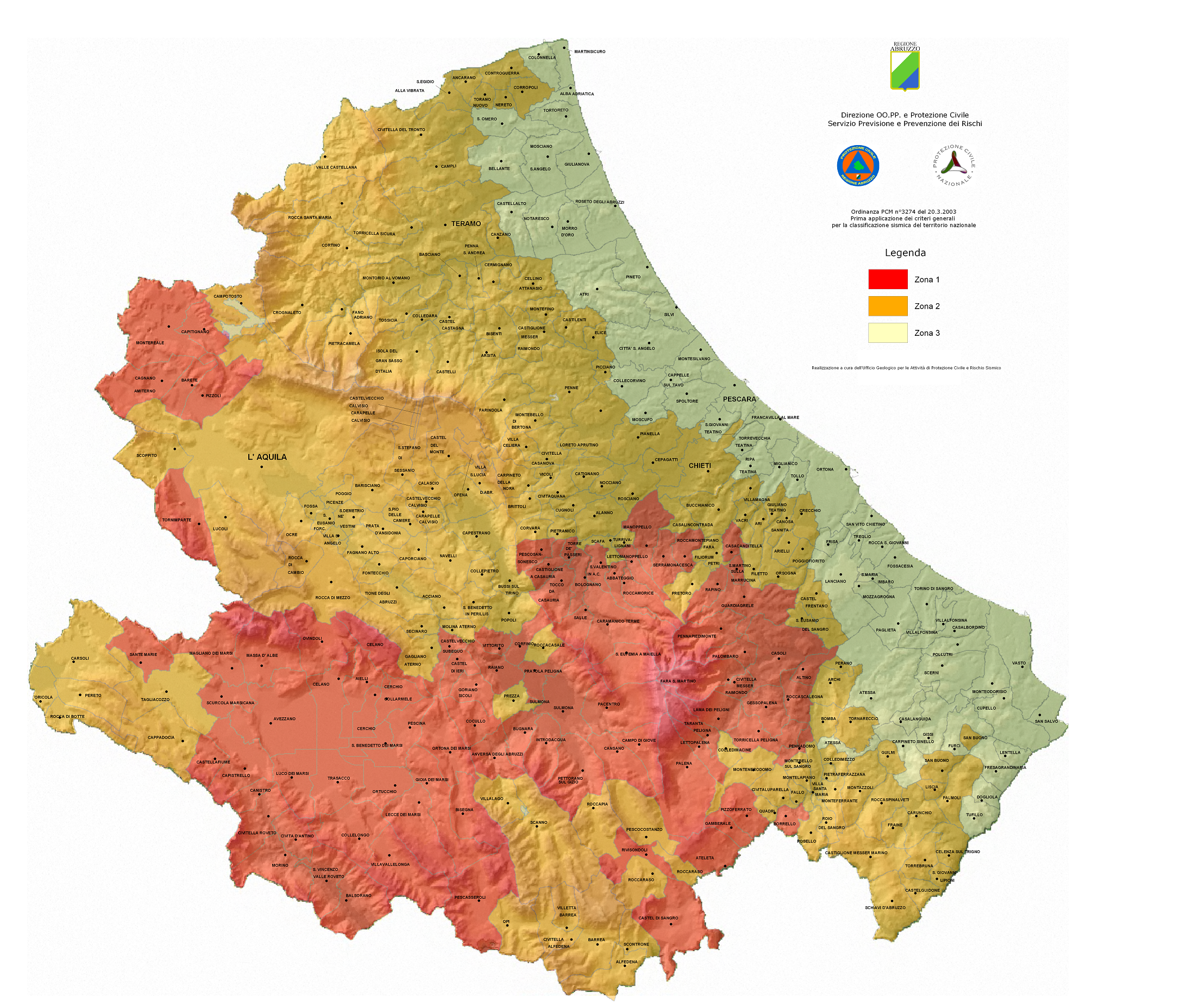} \\
{\scriptsize \textit{Fig.10: Map of Seismic Hazard (MPS) of Abruzzo. The acceleration values refer to a 475-year return (INGV2004). Legend: Zone 1 - Zone 2 - Zone 3}} {\scriptsize $^{\cite{INGV:2004}}$}\\ \\
The Aquila municipality, identified as a regional reference for this study, is characterized by a territory with medium-high seismicity and the elastic response spectrum (horizontal and vertical) of the seismic actions is highlighted in the following graph: \\ \\
\includegraphics[width=0.49\textwidth{}]{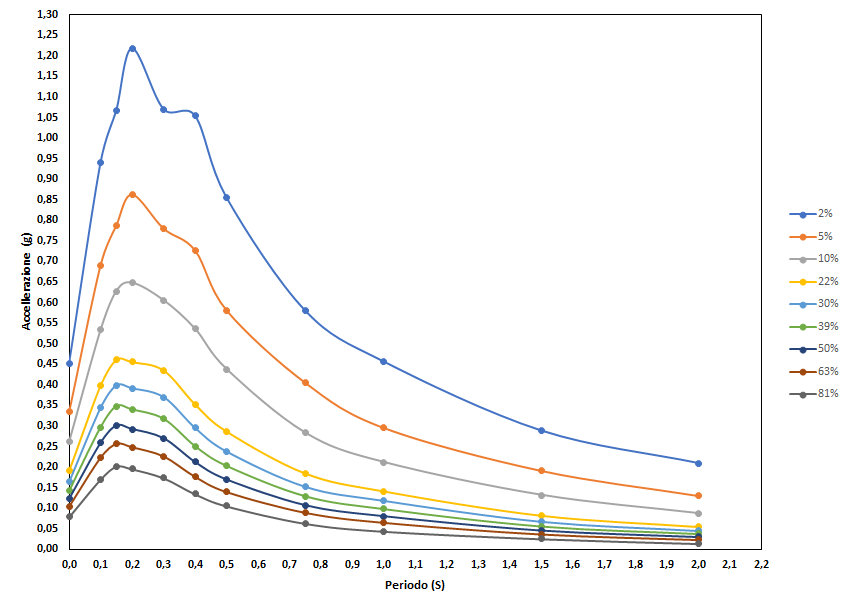} \\
{\scriptsize \textit{Fig.11: The spectra indicate the calculated shaking values for 11 spectral periods, ranging from 0 to 2 seconds. The PGA corresponds to the period of 0 seconds. The graph refers to the median estimates (50$^{\circ}$ percentile) proposed by the hazard model. The different spectra in the graph refer to different overflow probabilities (PoE) over 50 years.}} {\scriptsize $^{\cite{INGV:2004:ref2} \cite{Stucchi:ref1}}$}
\subsubsection{Umbria Region}
Seismicity in Umbria mainly occurred in the eastern and north-eastern sectors of the territory, reaching MCS intensity (Mercalli-Cancani-Sieberg) even equal to 10 (Norcia 1703, Gualdo Tadino 1751) and Magnitude (M$_{s}$ [Magnitude of surface waves]) over 6,5, with an earthquake frequency above the 7$^\circ$  MCS degree greater than or equal to 20 events per century in the last three centuries and overall for that period with 15 earthquakes having an intensity greater than or equal to the 8$^\circ$ MCS degree.\\ The distribution of earthquakes is a consequence of the arrangement of the seismogenic zones and their geological-structural characteristics. In general it can be stated that the area east of the Tiber River-Valle Umbra alignment is affected by a medium-high to high seismicity, while the one to the West is affected by a more modest degree of seismicity, from medium-low to medium.\\ \\
\includegraphics[width=0.49\textwidth{}]{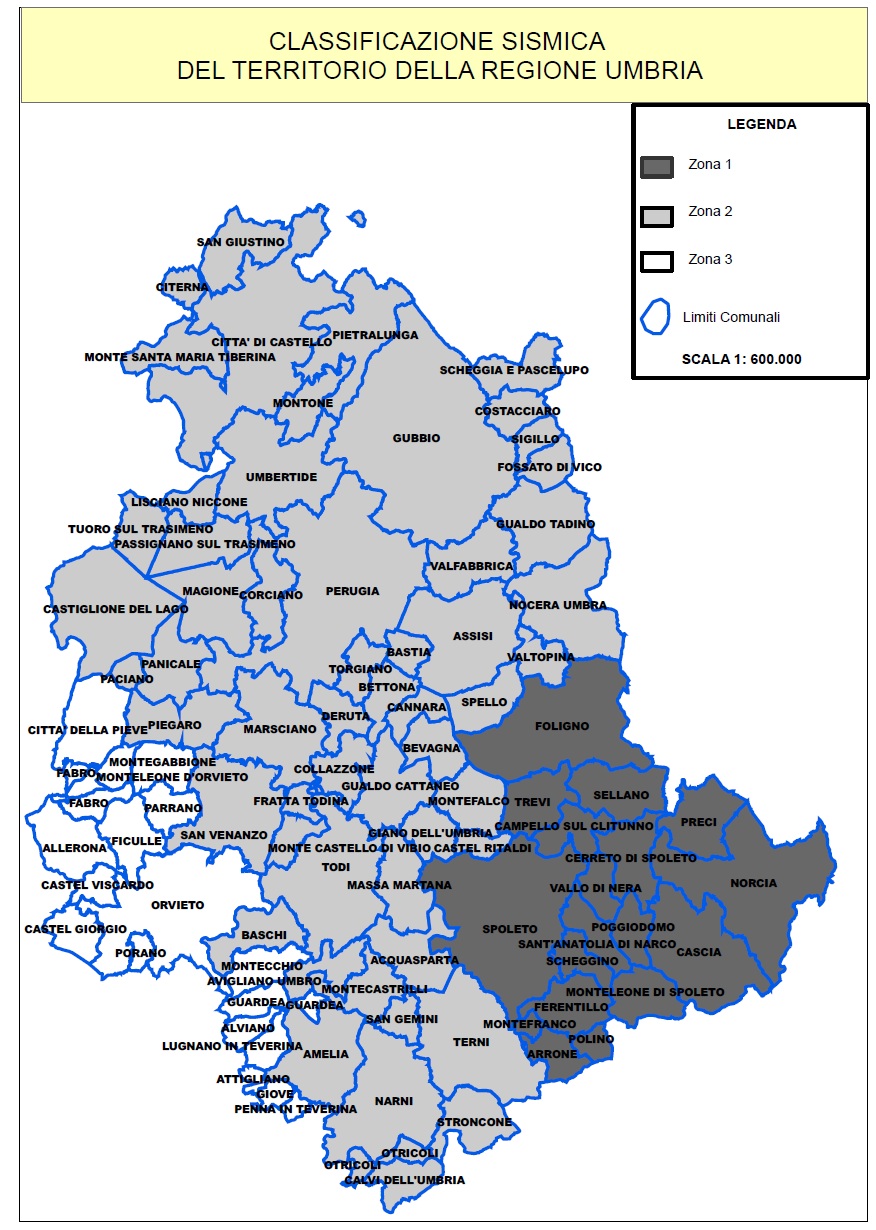} \\
{{\scriptsize \textit{Fig.12: Seismic Hazard Map (MPS) of Umbria. The acceleration values refer to a 475-year return (INGV2004). Legend:	Zone 1 - Zone 2 - Zone 3 - Municipal limits}}} {\scriptsize $^{ \cite{INGV:2004}}$}\\ \\
Sheet \textbf{324 "Foligno"} represents an area falling mostly in the province of Perugia and marginally in that of Macerata. This sheet has been used as a reference because it reports the data and the landslide event of \textit{"Serravalle di Chienti"}, located in the right margin of the sheet, better described in the "Reconstruction of Events" section.\\
From a geological point of view, the area is characterized by alluvial plains which meet with hilly and mountainous areas. \\In fact, the territory corresponds to a typical chain with folds and thrusts, in which wide ridges, anticlines and narrow synclinary valleys alternate.\\ This structure is locally interrupted by normal faults linked to the extension tectonics and recent uplift phenomena in the Apennine area. The faults also created new base levels for the forming hydrographic network and generated more regressive erosion. \\
The tectonic phase described briefly is still active and the structures related to it are responsible for the earthquakes that have affected and still affect the local area through micro-seismic events continuously present in the area.\\ Linked to the extensional structures of the axial zone of the Apennine chain are numerous earthquakes affecting the area highlighted in this study, and those affecting the areas further north of the Apennines, such as those of 1279, 1298 and 1328 described in this study in the section "Reconstruction of events", and highlighted in the elastic response spectrum (horizontal and vertical) of the seismic actions visible in the following graph relating to the city of Foligno: \\ \\
\includegraphics[width=0.49\textwidth{}]{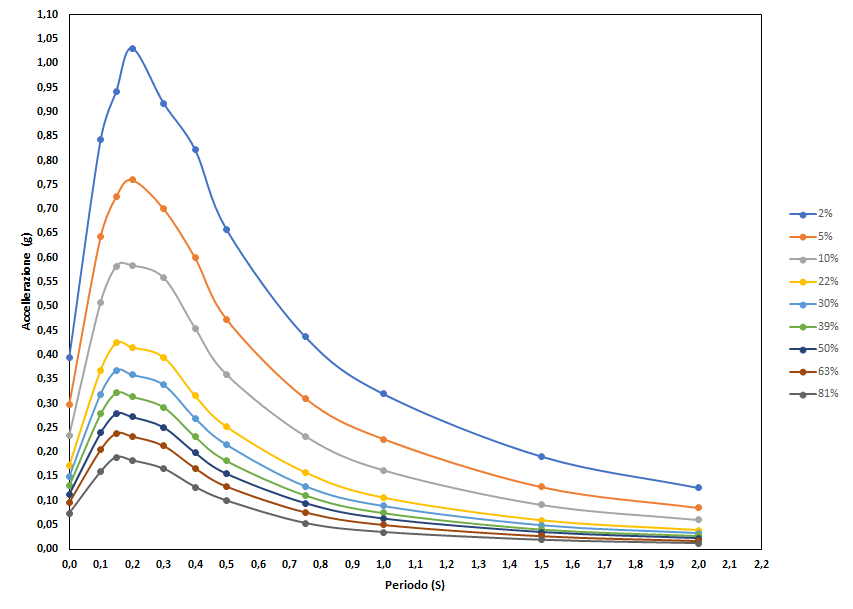} \\
{\scriptsize \textit{Fig.13: The spectra indicate the calculated shaking values for 11 spectral periods, ranging from 0 to 2 seconds. The PGA corresponds to the period of 0 seconds. The graph refers to the median estimates (50$^{\circ}$ percentile) proposed by the hazard model. The different spectra in the graph refer to different overflow probabilities (PoE) over 50 years.}} {\scriptsize $^{\cite{INGV:2004:ref2} \cite{Stucchi:ref1}}$}
\subsubsection{Marche Region}
\includegraphics[width=0.49\textwidth{}]{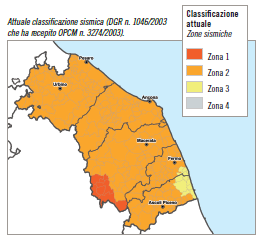} \\
{\scriptsize \textit{Fig.14: Seismic Hazard Map (MPS) of Marche. The acceleration values refer to a 475-year return (INGV2004).Legend:
Zone 1 - Zone 2 - Zone 3 - Zone 4}} {\scriptsize $^{\cite{INGV:2004}}$}\\ \\
The regional territory is characterized by a fairly uniform and medium-high level seismic hazard: this means that seismic activity is frequent and that earthquakes of high magnitude can occur, even destructive (although the latter, with a probability of occurrence lowest).\\ The acceleration values predicted by the seismic hazard model range from 0.15 to 0.25 g, with the highest values corresponding to the Umbria-Marche Apennine area.\\
The assignment of almost the entire region to seismic zone 2, with the exception of a small portion on the border with Abruzzo region in zone 1, and some municipalities on the coast in zone 3, is compatible with the expected ground shaking values ( PCM Ordinance No. 3519/2006).\\
The high seismic hazard values are determined by the presence of many seismically active structures and the seismic history of the Marche region, which reached its peak in the Apennines with the Cagliese earthquake of 1781 (magnitude $M_{W}$ 6.4) and on the coast with the Senigallia earthquake of 1950 ($M_{W}$ 5.8).\\
The municipality of Serravalle is characterized by a medium-high seismicity area; in fact, in table A attached to the Decree of 14/01/2008 of the Ministry of Infrastructures, the estimates of seismic hazard are highlighted and from these the elastic response spectrum (horizontal and vertical) of the seismic actions highlighted in the following graph was determined:: \\ \\
\includegraphics[width=0.49\textwidth{}]{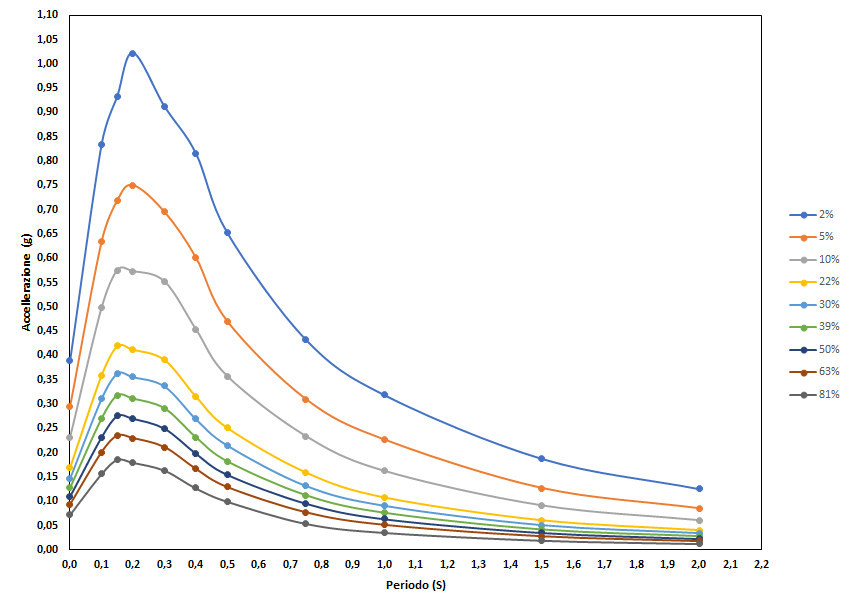} \\
{\scriptsize \textit{Fig.15: The spectra indicate the calculated shaking values for 11 spectral periods, ranging from 0 to 2 seconds. The PGA corresponds to the period of 0 seconds. The graph refers to the median estimates (50$^{\circ}$ percentile) proposed by the hazard model. The different spectra in the graph refer to different overflow probabilities (PoE) over 50 years.}} {\scriptsize $^{\cite{INGV:2004:ref2} \cite{Stucchi:ref1}}$} 
\subsection{The eruption of Samalas volcano in Indonesia in 1257}
The cores obtained in the last thirty years from the ice sheets of Antarctica and Greenland, have shown that around 1257 something truly shocking must have happened for the entire planet. From the sulfur deposits found between those layers of ice, it was deduced that a volcanic eruption had introduced into the atmosphere a quantity of sulphates at least eight times greater than that - already considerable - produced by the eruption of Krakatoa in 1883. The measured levels were even double those recorded on the occasion of the Tambora eruption in 1815 (an incredibly devastating event responsible for what the annals of history report as the year without summer).\\
From the stories of medieval European chronicles it emerges that something equally shocking must have happened around the year 1258. The critical environmental situation was also clearly evidenced by the growth rings of the trees corresponding to that distant period. Due to the reduced insolation, the incessant rains and the consequent floods, the crops were incredibly poor and the populations were wiped out by the famine and the inevitable diseases that accompanied it. \\By linking together all the available evidence, it emerged clearly that for the dramatic events that took place at the turn of 1257 the responsibility lay with a volcanic event. There was a well-founded suspicion that, given the almost uniform diffusion of volcanic sulphates in the two hemispheres, its location should be sought in the equatorial belt, but nothing more. However, a recent study that the volcanologist Franck Lavigne (University of Paris) and his collaborators published in PNAS finally seems to remove all residual doubts. Alongside the long-known geophysical evidence, the crucial testimony of some historical documents, known as Babad Lombok, which in ancient Javanese narrate on palm leaves the events of the kingdom that thrived on the Indonesian island of Lombok has emerged. From the stories emerges the devastating destruction of Pamatan, the capital of the ancient kingdom, carried out by the Samalas volcano, a structure that belongs to the volcanic complex of Mount Rinjani.\\Today on the summit of Mount Samalas there is a caldera of 6.5X8 Km, and on the slopes of the volcano there are imposing deposits of ash and traces of devastating pyroclastic flows. The reconstruction of that volcanic eruption that emerges from Lavigne's study leaves you speechless: the most conservative estimates speak of 35-40 Km$^{3}$ of material expelled from the volcano with a plume of ashes that pushed into the atmosphere up to an altitude of 43 Km. The devastation affected not only Mount Samalas and its immediate vicinity. Deposits of pumice up to 35 meters thick and attributable to the event, have in fact been identified in the surrounding islands even 25 km away from the volcano. According to volcanologists, the eruption was truly exceptional, characterized by a degree of explosiveness (VEI index) equal to 7, one of the most violent eruptions of the last 7000 years. Not only does Lavigne's study highlight how the radiocarbon dating of the materials is consistent with their location in the mid-thirteenth century, but also how the chemical analysis of the Samalas pumice deposits coincides with that of the materials found in the cores arctic and antarctic.\\ Finally the person responsible has been identified. Faced with such an eruptive power, it is not hard to believe that, as happened for that of Tambora, also following the eruption of Samalas, those devastating consequences for the entire planet, testified by medieval tales, could have triggered.
\section{Meteorology in Central Italy from 1000 to 1500}
It is not easy to summarize the climatic variations during the medieval period reported in this study, both because it encompasses a wide span of centuries, and because it is still difficult to enclose in a few lines an event, defined as the "period of the Medieval Heat" , which occurred in several phases, with ups and downs different from area to area, but characterized in any case by an average increase in temperatures which have reached much higher levels than today.\\ We begin by reporting that after a period of cold culminating in the fifth and sixth centuries, the climate is considerably mitigating on our continent. Furthermore, the presence of a much higher level of permanent snow than now, with the disappearance of glaciers and snowfields, allowed, during this period of medieval "Optimum", (which lasted approximately between 800 and 1200 AD ), the creation of new roads and new passes in the high mountains, and therefore greater possibilities of communication between the populations of the Alpine valleys, Italian and foreign, as well as the possibility of growing cereals up to relatively high altitudes.\\ The most favorable period of the medieval climate was certainly the XI century, 100 years during which in the European climate, only two harsh winters are reported, that of 1044, and above all that of 1010-11, which was really severe, as it was it formed ice around Iceland, froze the Bosphorus, and even froze the Nile in Cairo! \\ In this period "signs from the sky" that is an unusual series of northern lights visible as far as Mediterranean Europe, indicate a much higher solar activity than now, which also marked the beginning of the period of great heat; during this period there were also numerous droughts, due to the increased presence of the African Anti-cyclone.\\ 
The presence of numerous droughts, accompanied by invasions of African grasshoppers (in 873 they went up from Spain to Germany, in 1195 they reached Austria and Hungary, but sometimes even went as far as the Scandinavian countries), led to several medieval famines in the Mediterranean countries.\\ In Italy, further problems were caused by the rise in sea level, with the swamping of numerous coastal stretches, and the consequent proliferation of malaria in the lowland areas.\\ However, even in the warm twelfth century, some very harsh winters began to show up on our country from time to time, which began to intensify during the second half of the century. For example, the famous winter of 1162, when in Milan there were 40,000 deaths from cold and hunger, and almost all the crops were destroyed by the frost.\\ In 1200 the thermal drop almost suddenly arrived in Northern Europe; Iceland for example, which for 170 years had seen polar ice only once, was surrounded by them every year from 1197 to 1203 (the year during which these ice remained on the island also in the months of July and August). It has been hypothesized, recent discoveries on the shells of foraminifera show it, that 4 Centuries of heat had melted the ice to such an extent as to strongly weaken the Gulf Stream, with a loss of efficiency of at least 30\%.\\
In 1205 the Thames in London froze, in 1216 the Po froze; in the very famous winter of 1234 the Po, the Venetian Lagoon and the Thames froze from 25 December to 02 February, even with freezing of apple trees in England. The climate had changed, the glaciers suddenly advanced. The peat bog of the Fernau glacier in Tyrol was covered by ice between 1220 and 1350 (the alpine passes and the high altitude pastures covered with snow and ice), thus dating the post-medieval cold period, leaving imprinted in popular memory in the form of numerous legends, the change in the climate occurred.\\ The first years of the fourteenth century were very hard:
 \begin{itemize}
\item In 1303 the Arno waters froze, in January.
\item In 1305 it snowed until May in Central Italy, and the main rivers of Central and Northern Italy froze.
 \end{itemize}
The decade 1310-20 was then very hard:
after the umpteenth episodes of freezing of the Po river and the Thames in London, in the years 1310 and 1311, a period of continuous rains began; between 1315 and 1317, the entire European continent was hit, with one year, 1316, which was completely devoid of Summer and tormented by continuous rains, which prevented sowing on muddy fields.\\
The result was one of the worst famines of the entire Middle Ages, with tens of thousands of deaths across the continent. The cultivation of grapevine in England suffered a severe blow, so much so that it gradually disappeared. Numerous villages on the English west coast, which had thrived for 4 centuries, were washed away by storms, and never rebuilt.\\ From about 1350 onwards the climate mitigated again, but it no longer reached medieval levels. The harsh winters disappeared until 1408, with the sole exception of 1396 when the Tiber river, in the heart of Rome, could be crossed on foot for 15 days in a row.\\ \\ \includegraphics[width=1\linewidth]{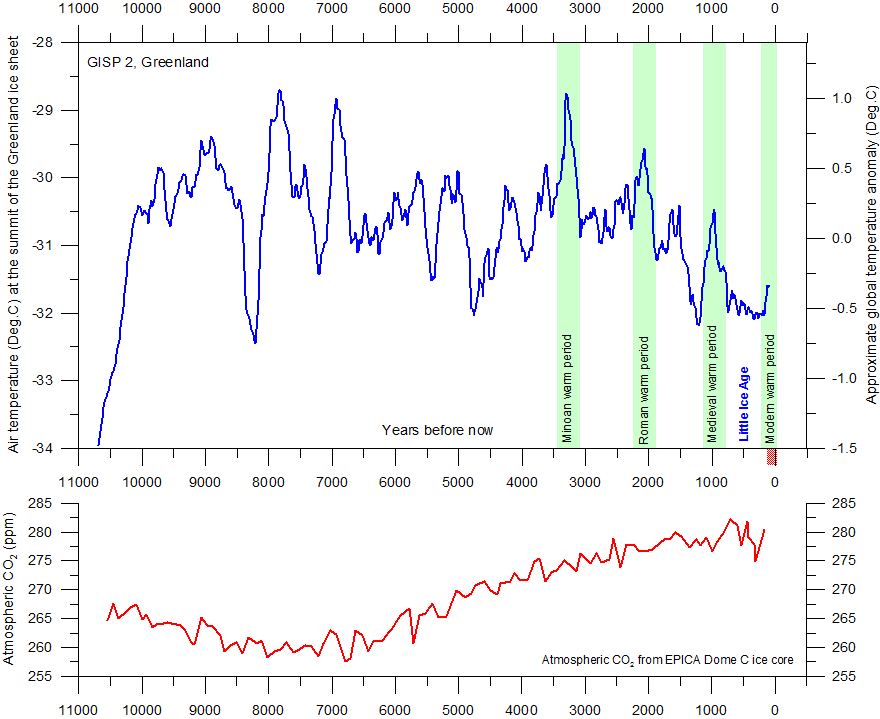}
{\scriptsize \textit{Fig.16: Temperature and values of CO$^{2}$ in the world. Highlighted, the impacting values are observed for the present study.}} {\scriptsize $^{\cite{graficoclima:ref1}}$} 
\section{Reconstruction of the events}
For the purpose of a correct exposition of the facts, the list of earthquakes that occurred in the period 1250-1350 (as present in the historical years recorded) is indicated in table 1, and below the events that have seen illustrious characters of history as witnesses present. at the event or on a subsequent visit (Pope Boniface VIII and Boccaccio).\\ I also indicate below some definitions of territories present in the historical period indicated and no longer known:\\
\textbf{\underline{Abruzzo Ultra}}: it was an administrative unit first of the Kingdom of Sicily and then of the Kingdom of Naples. Established in execution in 1273 by Charles I of Anjou with the "diploma of Alife", which formalized the division of the execution of Abruzzo, created by Emperor Frederico II, into two administrative districts, the Aprutium ultra flumen Piscariae and the Apriutium citra flumen Piscariae (Abruzzo on the other side of the Pescara river and Abruzzo on this side of the Pescara river). The capital was Aquila, and the boundaries of the execution included a large part of the current provinces of Aquila, Teramo and the part of the province of Pescara north of the homonymous river, in addition to the territories of the province of Rieti corresponding to the district of Cittaducale.{\scriptsize $^{\cite{citawikipedia}}$}\\
\textbf{\underline{Terra di Lavoro}}: historical-geographical region of southern Italy, divided between the current regions of Campania, Lazio and Molise.{\scriptsize $^{\cite{citawikipedia}}$}\\
\textbf{\underline{Contado di Molise}}: it was an administrative unit of the Kingdom of the Two Sicilies. Its territory mainly included the central part of present-day Molise and extended, depending on the time, in an area ranging from southern Abruzzo to northern Campania.{\scriptsize $^{\cite{citawikipedia}}$} 
\subsection{\normalsize Event of April 30, 1279}
The earthquake occurred at around 2 pm GMT \textit{"immediate post vespera"} according to the expressions of the time, affecting a large area of the Umbria-Marche region, taking shape according to the testimonies of the time, as an event that caused devastating damage to both the building structures and the environment. The most detailed source of information is the chronicle of the Franciscan Salimbene de Adam composed in the 80s of the 13th century, but we do not know if he was present, or if he was subsequently informed of the event. At the local level the event is reported by the Notary of Foligno Bonaventura di Benvenuto who lived from the end of 1200 to the beginning of 1300. The event perceived also in Rome but to a lesser extent, is reported in the \textit{"Continuatio Pontificum Italica II"} which tells the life of Pope Nicholas III (1277-1280). \\The affected villages were Camerino, Nocera Umbra, Cingoli, Fabriano, Matelica, San Severino Marche, Foligno, Spello and numerous unspecified local castles. The earthquake was felt in Rome and Montecassino south of the epicenter, and as far as Venice in the north. Subsequent strong tremors were felt for 14, 15 or 17 days. \\In Camerino two of the four areas into which the city was divided were severely affected; \textit{"La Chronica S.Petri Erfordensis"} reports the destruction of all the towers and town houses, and indicates the number of dead as 1.000. \textit{"Gli Annales Polonorum"} reported that fifty nuns died in a circestian convent. Historian Camillo Lilii (1649-1652) reports the collapse of the bell tower of Santa Maria, the tower of San Giacomo and a monastery. \\The Castello di Serravalle locality indicated in the medieval annals with "Serravalle" or "Castrum", was affected by the event due to concomitances both from a building and environmental point of view, as a landslide broke away from the mountains, burying all the citizens (about 500 people). \\Nocera Umbria was heavily hit and over half of the city was destroyed; \\the monastery of the main church collapsed together with the surrounding buildings used as the residence of the canons. \\Cagli is included among the places that suffered extensive damage, such as Cingoli, Fabriano, Foligno, Matelica, Spello and San Severino Marche.\\ In Cerreto di Spoleto, the population gathered outside the walls of the castle holding council, both because the earthquake had caused serious damage to the buildings, and also as a precaution since the city had been hit by a previous earthquake in 1277.{\scriptsize $^{ \cite{pubblicazione:ref1}\cite{pubblicazione:ref2}}$}
\subsubsection{\normalsize The landslide in the Serravalle castle area}
The geological/historical analysis has led to identify that "Serravalle" indicated in the historical sources, was close to or coincided with the locality "Castello di Serravalle", exposing that the data to the buildings were caused by the landslide that broke away from the mountain above, which brought about a temporary change in the local water regime, a swamping of the area, and a deviation of the river.\\ \\
\includegraphics[width=1\linewidth]{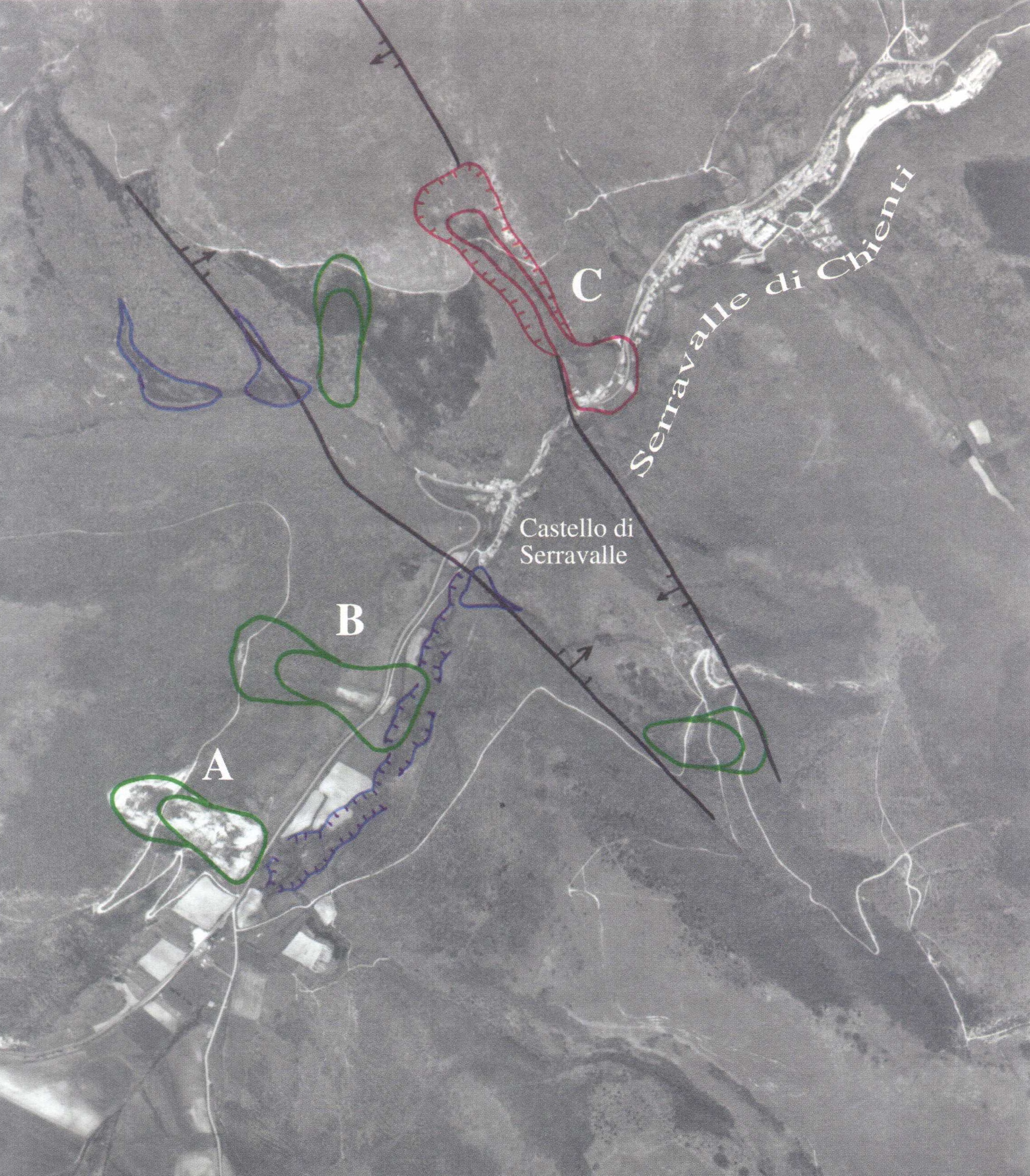} 
{\scriptsize Fig.17: Location of the historical landslides of the upper valley of the Chienti river, interesting the town of Serravalle di Chienti presented in this study.}{\scriptsize $^{\cite{pubblicazione:ref1}\cite{pubblicazione:ref2}}$} \\ \\
The geological analysis led to the identification of three landslide events described below: 
\begin{itemize}
	\item the first landslide (A) located near the Colfiorito quarry is recent and is related to the activities of a quarry which took place in 1988-1989;
	\item the second landslide (B) located a little further north on the slopes of Mount Faeto, has morphological and lithological characters similar to the previous one. These are materials that reached the valley floor temporarily occluding the Chienti river;
	\item the third landslide (C) located further north of the town of Castello di Serravalle. Unlike the others, the landslide also affected the rocky substratum, and extends from 950 meters to 560 meters. The movement, which involved 1 million m$^{3}$ of material, reaching the valley floor blocked the flow of the Chienti river, causing temporary swamping of the area upstream of the landslide.\\	The morphological characteristics, the type of movement due to the collapse caused by a fault movement present in the area, and the shape of the deposit present at the foot of the slope, tend to make us believe that the landslide movement coincides with the seismic event of 1279. {\scriptsize $^{ \cite{pubblicazione:ref1}\cite{pubblicazione:ref2}}$}  
\end{itemize}
\subsection{\normalsize Event of May 01, 1279}
In the night between April 30 and 1$^\circ$ May 1279 a strong earthquake struck southern Romagna on the border with Tuscany, in the current provinces of Forlì and Florence: it can be argued that this earthquake is clearly distinct from that of the Umbrian-Marche Apennines on the afternoon of 30 April.\\
The earthquake struck some towns and fortresses of the high Romagna Apennines. The collapse of many fortified villages and their walls in the mountainous areas between Bologna and Florence (or between Romagna and Tuscany) is generically mentioned, without a precise location.\\
Specific effects are reported for only some locations.\\ The tower and the castle of Fantella collapsed (located between Portico di Romagna and Galeata in the valley of the river Rabbi, province of Forlì), in which some people died; the castle was never rebuilt and in a fiscal census in the last quarter of the 14th century, promoted by the cardinal legate Anglico for the whole of Romagna, Fantella is classified as a "villa", that is, as a center that is no longer fortified. The ruins of the castle are located in Castellaccio (near today's parish church of Santa Maria di Fantella).\\ In Castiglionchio (today Rocca di Castiglione, near Marradi, in the province of Florence) the fortress collapsed and there were many deaths. \\ The monastery of S. Ellero di Galeata (province of Forlì) was destroyed. {\scriptsize $^{\cite{pubblicazione:ref1}\cite{pubblicazione:ref2}}$}\\ \\ \\
\subsection{\normalsize Event of December 01, 1298}
The earthquake defined as "del Reatino" mainly affected the areas of Rieti and Spoleto.\\The main quake probably occurred on December 1, 1298, although other sources report dates as different as November 20 or November 10. It had its epicenter in the Reatini mountains between the centers of Leonessa (RI) and Polino (TR). It was preceded on November 30 by a shock of lesser intensity (magnitude 4.4) with its epicenter in Rieti and was followed by strong repetitions that lasted for about six months; according to the chronicles of the time, the aftershocks were so strong that they knocked people to the ground.\\
Leonessa was one of the most severely affected cities: numerous buildings collapsed, including the church and convent of San Francesco (now home to the Civic Museum). Very serious consequences were also had in Poggio Bustone, where damage equal to the IX-X$^\circ$ degree of the Mercalli scale was recorded: according to the chronicles of the time, the town "was ruined from its foundations" and 150 people died.
The castle of Vetranola, near Monteleone di Spoleto, was completely razed to the ground and had to be abandoned. In 1302 the village was rebuilt, but it was decided to rebuild it in a different position, in the hope of making it less vulnerable to earthquakes (the hope was in vain: the town will be destroyed again, this time for good, in the earthquake of 1328).\\
In Rieti and Spoleto the earthquake caused serious damage, equal to VIII$^\circ$ degree on the Mercalli scale. In Rieti many buildings collapsed and others were seriously damaged, in particular the Papal Palace and the towers. The signs of the repairs carried out following the earthquake are still visible today on Palazzo Potenziani.\\The earthquake also caused damage to Subiaco and was felt in the Marches, in Verona, in Forlì and in Rome. In Todi for fear of the earthquake, most of the population fled the cities and took refuge in the countryside.\\
The earthquake caused a large number of victims globally, although the total number is not known.\\We have news of the earthquake thanks to the reports of many contemporary and non-contemporary historians: the event is mentioned by Giovanni Villani, Bernardo Gui, Platina, Panvinio and Chacón; in particular it is mentioned in many historical sources by the biographers of the popes due to the involvement of Pope Boniface VIII, who at that time was in Rieti (the city had been the papal seat several times during the 13th century). The chronicles of the time report that at the moment of the quake, the pontiff was celebrating mass in the Cathedral and that he was so frightened that he fled to the countryside, with his vestments still on. The pope was forced to take refuge in a camp tent set up in the cloister of the Dominican convent.\\
The earthquake, as well as the simultaneous appearance of a large comet, was considered by the public opinion of the time a divine punishment against Boniface VIII, who in those years was waging a bloody struggle for power against the Colonna, and a sign of future personal misfortunes.\\Following the earthquake, the pope himself promoted a series of interventions for the reconstruction and consolidation of the damaged buildings in Rieti. Among these we remember the Arch of Boniface VIII, built to consolidate the nearby Papal Palace. {\scriptsize $^{\cite{pubblicazione:ref1}\cite{pubblicazione:ref2}}$}
\subsection{\normalsize Event of December 03, 1315}
The earthquake of 3 December 1315 also known as the "Aquila earthquake" was a seismic event that occurred in the middle Aterno valley, in Abruzzo. The earthquake had a moment magnitude of 5.6 and an intensity between the VIII and XI$^\circ$ degree of the Mercalli scale. This is the first major seismic event to hit the city of Aquila, built in the middle of the previous century, and caused numerous damage to the artistic and architectural heritage.\\
As reported by direct historical sources, including the Chronicle of Aquila in rhyme of Buccio di Ranallo, the seismic swarm began in February 1315, but the main shock occurred on Wednesday 3 December. At the moment the seismic origin of the event is particularly debated, which today is identified in the middle Aterno valley, in the center of the Aquila basin, while other previous studies located the epicenter about 60 km south-east of L'Aquila, in the northern part of the peligna basin, near Sulmona. The estimate of the power of the event also varied over time between the values 5.1 and 6.0 and is now calculated in 5.5$\sim$5.6 moment magnitude, with an intensity between VIII and XI$^\circ$ on the Mercalli scale.\\ The seismic sequence continued for about 4 weeks and if the replicas continued for several months also in 1316, and for a long time the inhabitants took refuge in temporary barracks built in the fields around the city.\\
The earthquake is of particular importance as mentioned above, because it was the first to hit L'Aquila, having occurred about sixty years after its foundation. In fact, at the date of the earthquake, the city had just completed its first expansion and, under the guidance of Captain Lucchesino from Florence, the construction of the city walls was being prepared.\\
The news on damages and casualties, however, are rather meager. \\ The Cathedral certainly suffered very serious damage, thus starting a troubled history that caused its actual completion only in 1928; the churches of San Francesco a Palazzo and San Tommaso di Machilone were also affected, which have now disappeared. \\ Historical accounts report that the city administration announced the reconstruction of the Machilone church as a vote for the earthquake, but the promise was not fulfilled. With the exception of this last monument, the reconstruction was immediate, so much so that the new city walls were already completed the following year.{\scriptsize $^{\cite{pubblicazione:ref1}\cite{pubblicazione:ref2}}$}
\subsection{\normalsize Event of December 04, 1328}
The earthquake occurred at dawn around 6:15 am on December 4th; the epicenter was near Roccanolfi, about 9 km north of Norcia, on the western side of the Sibillini mountains. The event recorded a moment magnitude of 6.4 and an intensity equal to the X$^\circ$ degree of the Mercalli scale.\\
In Norcia there was the collapse of numerous palaces, churches and towers, as well as part of the city walls. \\Even more serious damage was recorded in Montesanto and especially in Preci, where the collapses caused the death of all the inhabitants of the town. \\Castel San Giovanni near Cascia, Cerreto di Spoleto, Monte San Martino and Visso were seriously damaged.\\ Serious collapses in homes were also recorded in Spoleto. The earthquake was also felt in Foligno and Rome.\\
The news about the earthquake reached us through the direct testimony of the time of the notary Bonaventura di Benvenuto and included in his "Cronaca di Foligno", and a memorial written by the Jewish community of Ripatransone, in the Ascoli area; these sources made it possible to correct the exact date of the earthquake. \\The event is also referred to in the \textit{"Annales Arretinorum Minores"} by Giovanni Villani and in \textit{"Chronicon Mutinense"} of Giovanni da Bazzano, as well as in \textit{"Annali e cronaca di Perugia in volgare dal 1191 al 1336"} by Mariano del Moro and in the Perugian chronicle known by the name of \textit{"Diario del Graziani"}. \\The individual sources differ on the estimate of the victims which varies from 2,000 to 5,000 units.\\ \textbf{A little curiosity}: The memory of the earthquake survived for a long time among the inhabitants of the Valnerina; \\a memory of 1599 tells us how the population of Norcia celebrated the feast of Santa Barbara by commemorating the victims of the earthquake that occurred almost three centuries earlier. {\scriptsize $^{\cite{pubblicazione:ref1}\cite{pubblicazione:ref2}}$} \\
\subsection{\normalsize Event of September 09, 1349}
The event that affected the central-southern Apennines in 1349 was an earthquake, or more likely a set of seismic events, which occurred on 9 September 1349.\\ The earthquake was strongly felt in most of the State of the Church (except in Romagna) and the Kingdom of Naples (except in Calabria), while producing its maximum effects in the territory corresponding to the current regions of Abruzzo, Lazio and Molise. \\The exact epicentral location and the temporal sequence of events remain somewhat uncertain, but the entire seismic crisis resulted in about 2.500 victims.\\
Historical sources report that the sequence began on January 22, 1349 with a modest event that had its epicenter in Sannio, near Isernia, as recounted by Sant'Antonino di Firenze in his \textit{"Chronicon"}.\\It is also pointed out that the second very violent pandemic of \textbf{"black Plague"} which soon spread throughout Europe, while southern Italy suffered a heavy economic and social situation due to famine and civil wars.\\
It is not yet historically ascertained whether on 9 September several distinct earthquakes occurred almost simultaneously, or whether it was a single major seismic event with widespread effects throughout central-southern Italy. This second hypothesis is considered implausible while it is known that the seismic sequences in the Apennines can culminate in several destructive episodes within a short period of time. What makes the 1349 earthquake unique is that the seismic source is different in both (or four) events.\\
On the same day, therefore, "the highest concentration of disastrous earthquakes in Italian history" would have occurred, which, combined with the second pandemic of black plague, configured 1349 as \textbf{"annus horribilis"}. Table 2 at the end of the work specifies these seismic sequences.{\scriptsize $^{\cite{pubblicazione:ref1}\cite{pubblicazione:ref2}}$}\\ \\
By extension and intensity, the seismic sequence produced very serious damage in numerous centers of Abruzzo Ultra, of the Terra di Lavoro, of the Contado of Molise and of the Papal State, also affecting Rome and Naples. In this regard, an authoritative testimony is that of Francesco Petrarch who, on the occasion of the Jubilee of 1350, visited Rome finding it strongly tested by the damage caused by the earthquake; Petrarch notes that many civil buildings and many churches had collapsed due to the earthquake: he wrote in fact:\\ \\ "«\textit{Rome has been shaken by an unusual tremor, so badly that nothing like this has ever happened since its foundation, which dates back over two thousand years.}» \\ \\News of damage in the city also comes to us from the historian Matteo Villani who told how the earthquakes \textit{«they knocked down the bell tower of the great church of San Paolo, with part of the noble tower of the Militias, and the tower of the Count»}.\\ Giovanni Villani and above all Buccio di Ranallo, note in their Chronicles the devastation that the earthquake caused to l'Aquila, which was already damaged by the earthquake of 1315; there were many collapses, with Count Lalle Camponeschi who ordered to pile up the rubble near Porta Leone, thus preventing the frightened inhabitants from leaving the city, which alone counted about 800 victims.\\
According to a memory preserved in the abbey of Montecassino, the seismic event severely hit the entire Kingdom of Naples. Isernia, Venafro and Montecassino also suffered serious damage. \\ In Telese Terme nel Sannio, the repeated seismic tremors shook the ground, favoring the birth of mofetes, the creation of a lake and the emanation of sulphurous vapors that made the air unbreathable and caused the abandonment of the town. \\ In Alatri the earthquake caused the collapse of the central body of Palazzo Gottifredo. 
\subsection{\normalsize Events of December 25, 1352 and January 01, 1353}
On 25 December 1352 and 1 January 1353 two devastating earthquakes struck the upper Tiber valley, affecting in particular the municipalities of Borgo San Sepolcro (now Sansepolcro) and Città di Castello. The sources of this earthquake have been analyzed and discussed by Boschi et al. (1995).\\ The first earthquake occurred towards sunset (about 4:30 pm UT) and caused the collapse of some buildings in Sansepolcro. The foundation stones of the walls that rose above the outermost moat, albeit at a depth of over 4 meters, were thrown out of the ground. \\ The effects in Città di Castello were less serious, but the tower of the castle collapsed and the castle of Elci was destroyed, located on the border between the territories of Sansepolcro and that of Arezzo, where today the Torre dell'Elci stands.\\
The second earthquake caused the complete collapse of the walls and of almost all those buildings in Sansepolcro that had not collapsed in the earthquake of 25 December.\\ The bell tower of the abbey where the municipal documents were kept collapsed.
Many were the victims: 500 were the dead in Sansepolcro in the first earthquake, and the total number of dead was over 2.000 (or 3.000).\\
As far as contemporary chronicles are concerned, the main sources are the\textit{"Cronica di Matteo Villani"} (edited by Porta 1995), in which the effects of the two Sansepolcro earthquakes and the reconstruction of those buildings that had been destroyed are described in detail, and the \textit{"Liber gestorum in Lombardia"} by the Lombard notary Pietro Azario (edited by Cognasso 1925-39).\\ The latter records the effects of the earthquake in Città di Castello, of which the author claims to have been a witness.{\scriptsize $^{\cite{pubblicazione:ref1}\cite{pubblicazione:ref2}}$} \\ \\
\section{Epidemics in the study period}
The historical period described saw the European populations of the time affected both by extreme geological-climatic events and by the Plague (the most important epidemic disease of the past, which produced dramatic effects in the second half of the 1300s), a scourge that was called "\textbf{Black Death}", by defining 1349 as \textbf{"annus orribilis"}. \\The pandemic that hit Europe in the years 1347-50 was undoubtedly the most devastating ever in human history. \\In the map at the end of the article (fig.27) the propagation of the plague in Europe in the period 1347-1350 is visible.
\subsection{The plague from 1347 to 1351} 
As we have seen, periodic cycles of climate change have led to recurring outbreaks of plague. The first of 1257, triggered by the eruption of the Samalas volcano in Indonesia$^{\cite{lavigne:ref1}}$\normalsize, confirmed by the discovery of over 10 thousand skeletons dating back to the medieval period, found in the eastern part of London in the nineties and precisely at Spitalfields market during archaeological excavations. \\For the authors of the discovery it would be only the tip of the iceberg: it is believed that at least 15.000 people were buried in those mass graves, about a third of the London population of the time. \\A terrifying massacre that was initially attributed to the terrible plague known as the "black death", a contagion that plagued Europe at the beginning of the XIV$^\circ$ century.\\ However, this attribution was disavowed by radiocarbon dating, which clearly indicated that those deaths were from a few decades prior. \\
Originating from Central Asia, it reduced the world population from 450 million to 350-375 million. In Europe, the demographic upheaval was enormous, the population was reduced by a third, which went from 75 to 50 million. The "Black Death" contributed to the destruction of the medieval feudal system. \\In the years 1347 and 1348, when the plague spread throughout Europe under the name of the Black Death, society was completely unprepared to face it. \\The plague had long been absent from Western Europe and although some sporadic scholars knew that bubonic symptoms had been observed during Justinian's plague in the VI$^\circ$ century, it was mostly a new disease. It was new, not only in its nature, but also in the mortality rate it caused: about a third of the European population, and perhaps more in populous cities. City after city, chroniclers recorded the accumulation of corpses in homes, on the streets and in public places. Most ended up in huge pits, where they buried themselves quickly and without ceremony.\\ In addition to the dramatic demographic and public health impact, the plague paralyzed the political, social, administrative and commercial aspects of the city. The legislative instruments were lacking, the staff of any institution. Fear choked the hearts and minds of the citizens before the plague even arrived. The impending catastrophe strengthened the belief of many that the prophecies of the Apocalypse were coming true. Those who showed signs of the disease could no longer hope for compassion or for the help of others, including family members.{\scriptsize $^{\cite{pubblicazione:ref1}\cite{pubblicazione:ref2}}$}\\ \normalsize 
The pandemic of plague of 1347-1351 was admirably described by Giovanni Boccaccio who in his introduction to \textbf{"Decamerone"} che reminds us that the scourge was considered a divine punishment for the sins of humanity or the effect of an astral conjunction, but he was well aware that the scourge came from the east. He describes the symptoms of the disease precisely:\\ “\textit{…in the initiation of it, males and females alike were born either in the blood vessel or under the fingertips, some of which grew like a common apple, others like an egg, and some more and some less, which the vulgar names named gavoccioli…}”. {\scriptsize $^{\cite{Pasini:ref1}}$} \\ \\
He was aware that the medical intervention was of no effect: \\“\textit{After which infirmity, no medical advice, no virtue of medicine, it seemed that it was worth or profited…}”. {\scriptsize $^{\cite{Pasini:ref1}}$}\\ \\ He was aware that the infection also took place through contact with clothing or objects:\\ “\textit{…And later still he had a bad thing: because not only did talking and using with the sick give healthy infirmities or causes of common death, but touching the clothes or anything else that those sick had touched or used seemed to with him that such infirmity in the touchor carry…}”.{\scriptsize $^{\cite{Pasini:ref1}}$}\\ \\ \normalsize
Boccaccio recalls that in the face of the spread of the disease, the reactions of the population were different:\\“\textit{…And there were some who warned them that living moderately and guarding oneself from any superfluousness would have to resist very much accidentally: and having made their brigade, separated from each other they lived, and in those houses they gathered and locked themselves up where no one sick was and to live. better, very diligent foods and excellent wines, very temperate using and every lust fleeing, without letting anyone speak to anyone, or wanting outside, about death or the sick, any news to hear, with sounds and with those pleasures that they could have lived. Others, in contrast to oppinion traits, affirmed drinking a lot and enjoying and going around singing around and amusing and satisfying everything to the appetite that one could and of what happened laughing and mocking being a very certain medicine to so much evil: and as they said, the they put in work at their power, day and night now to that tavern now to that other going, drinking without way and without measure, and much more this for the other houses doing, only that things they could hear you that they came in degree or in pleasure…}” {\scriptsize $^{\cite{Pasini:ref1}}$} \\ \\ \normalsize The threat of the scourge was so strong and the terror of death was such as to upset not only the feelings of solidarity, but also the most sacred affections.:\\“\textit{…And let it be that the one citizen disgusted the other, and almost no one nearby had the other care, and the relatives together rarely or never visited each other and from afar; this tribulation entered the breasts of men and women with such fright that one brother abandoned the other, and the uncle the nephew, and the sister the brother, and often the woman her husband; and, what is greater and almost not credible, the fathers and the mothers the children…}” {\scriptsize $^{\cite{Pasini:ref1}}$}  
\section{Conclusions}
The temporal events narrated in this work lead to confirmation of how climate and geology impact local populations at any global latitude. \\
With the detailed descriptions of the numerous chemical-physical processes present in the aforementioned article, it is possible to affirm, perhaps even with absolute certainty, that a hypothetical eruption of a supervolcano (or more eruptions), if sufficiently violent and lasting, could cause a global cooling or even an ice age, but in the same way with equal confidence it can be assumed that it would still be a transitory phase, because as already happened in the past, nature and our planet will always find a way to self-regulate and ensure that the life survives once again.\\As regards the present study, it can also be hypothesized that the phenomenon of \textit{"lack of contracts"} by the Popes in certain periods at the turn of adverse geological and climatic events, both followed a \textit{"flourish again"} of the creation, renovation and embellishment of churches, monasteries and religious buildings. These are events of this type that have led, after their occurrence, to a period of greater momentum in the culture of beauty, leading to the production and creation of brand new architectural, pictorial and fresco.  
\section{Acknowledgments}
\normalsize We thank Dr. Fiorella Cerrone and teacher Roberto Rappa for their useful suggestions and their support in order to make the paper more complete. \\Constructive comments are very much appreciated. 
\end{multicols}
\newpage
\normalsize {\bf {Table 1: Number of Earthquake in center Italy from 1250 to 1350.}} \\ \\
\begin{tabular}{|c|c|c|}
	\hline
	\normalsize \textbf{Date} &\normalsize \textbf{Location} &\normalsize \textbf{Magnitudo} \\
	\hline
	\normalsize 30/04/1279 & \normalsize Umbrian-Marches Apennines & \normalsize 6,5 Richter/IX Mercalli. \cr
	\hline
	\normalsize 01/05/1279 & \normalsize Tuscan-Romagna Apennines & \normalsize 5,9 Richter/VIII-IX Mercalli. \cr
	\hline
	\normalsize 01/09/1298 & \normalsize Reatini Mountains & \normalsize 6,3 Richter/X Mercalli. \cr
	\hline
	\normalsize 03/12/1315 & \normalsize L'Aquila & \normalsize 5,6 Richter/IX Mercalli. \cr
	\hline
	\normalsize 04/12/1328 & \normalsize Norcia & \normalsize 6,4 Richter/X Mercalli. \cr
	\hline	
	\normalsize 09/09/1349 & \normalsize Central-southern Apennines & \normalsize 6,8 Richter/X Mercalli. \cr
	\hline
	\normalsize 25/12/1352 & \normalsize San Sepolcro on the Umbrian border & \normalsize two shakes of 5,6 Richter/VIII Mercalli. \cr
	\hline
	\normalsize 01/01/1353 & \normalsize San Sepolcro on the Umbrian border & \normalsize 5,8 Richter/IX Mercalli.\cr
	\hline
\end{tabular}
 \\ \\ \\
\normalsize {\bf {Table 2: List of seismic events, certain or hypothesized, relating to 9/9/1349 and listed from north to south}}\\ \\
\begin{tabular}{|c|c|c|l|}
	\hline
	\normalsize \textbf{Epicenter/Town} &\normalsize \textbf{Coordinate} &\normalsize \textbf{Mw} &\normalsize \textbf{Note}\\
	\hline
	\normalsize Montefiascone & \normalsize $42^\circ 31' 48$'' N $11^\circ 58' 12$'' E & \normalsize - & \normalsize  Hypothesized in northern Lazio, on the border between\cr 
	& & &the Papal State and the Grand Duchy of Tuscany.\cr
	\normalsize Fiamignano & \normalsize $42^\circ 16' 12$'' N $13^\circ 07' 12$'' E & \normalsize 6.3 & \normalsize  Cicolano area. Serious damage to l'Aquila with about \cr
	& & & 800 victims.\cr	
	\normalsize Sulmona & \normalsize $42^\circ 01' 12$'' N $13^\circ 56' 24$'' E & \normalsize - & \normalsize Hypothesized in the Peligna basin, on the border between\cr
	& & & Abruzzo Ultra and Abruzzo Citra.\cr	
	\normalsize  Acquafondata & \normalsize $41^\circ 33' 00$'' N $13^\circ 56' 24$'' E & \normalsize - & \normalsize  Border between Terra di Lavoro and Papal State.\cr
	& & & Effects propagated over a large area.\cr
	\hline	
\end{tabular} 
 \\ \\ \\
\includegraphics[width=1.35\linewidth]{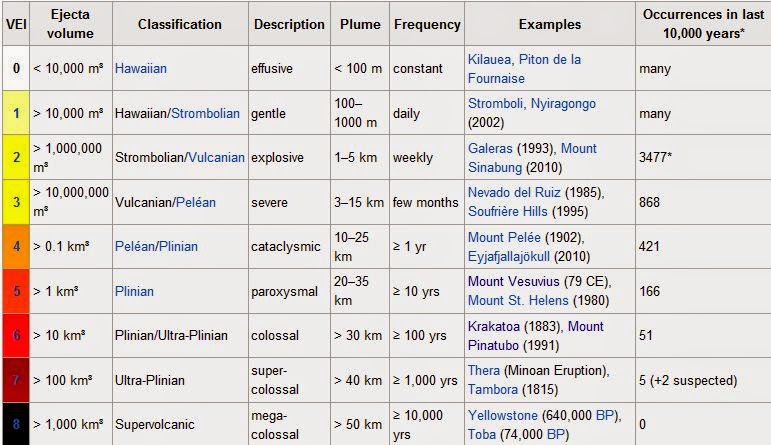}
\normalsize {\bf {Fig.18: VEI - Volcanic Explosivity Index - Source Wikipedia}} \\
\includegraphics[width=1\linewidth]{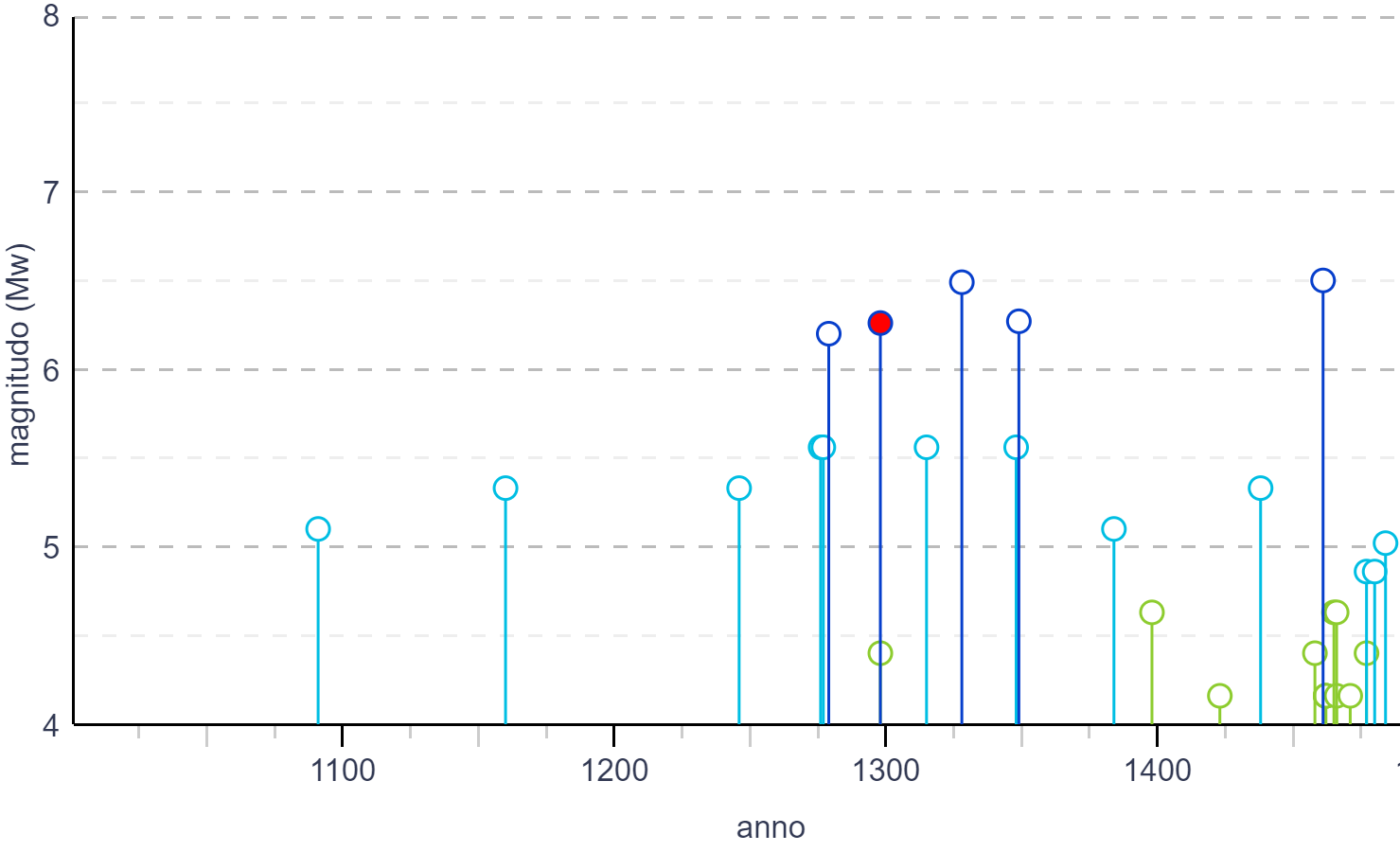}
\normalsize {Fig.19: Seismicity felt in Rome in the period between 1000 and 1550. The red dot identifies the event of December 01, 1298 attended by Pope Boniface VIII, reported in the section "Reconstruction of the Events".}\\
\\ \\
\includegraphics[width=1\linewidth]{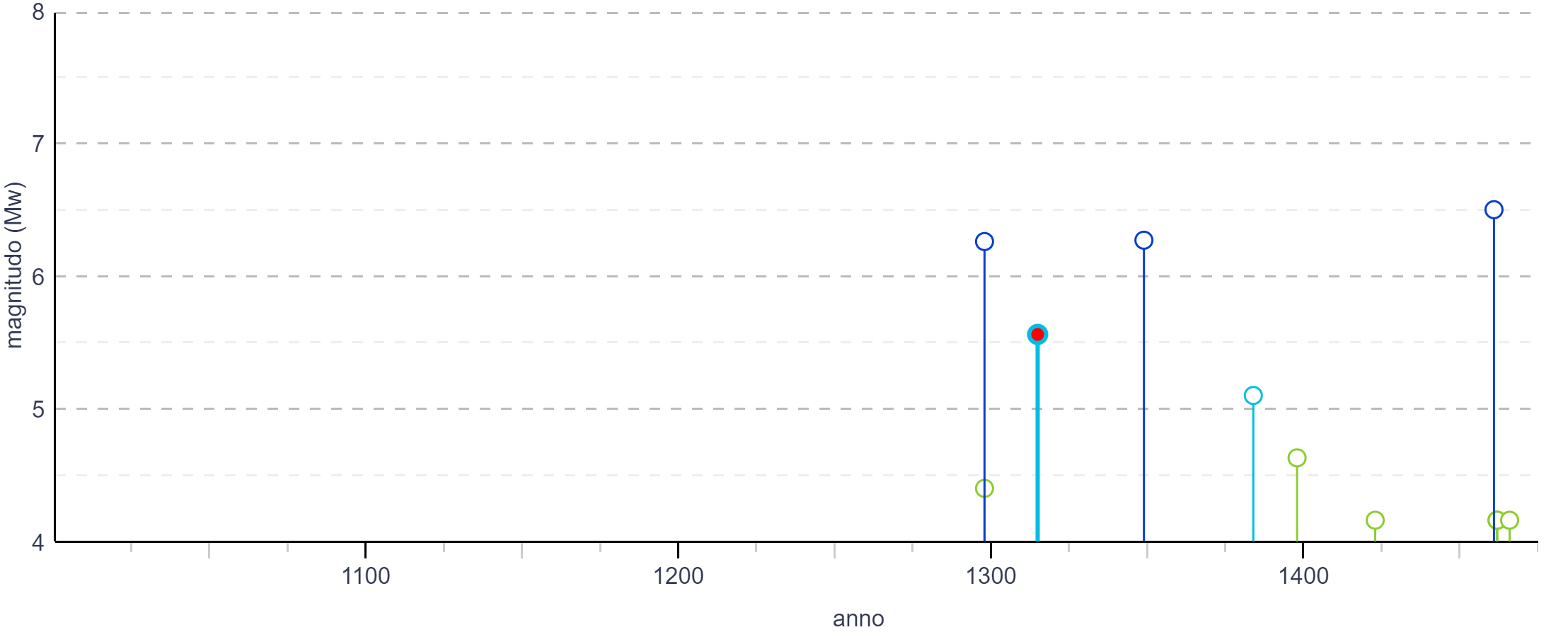}
\normalsize {Fig.20: Seismicity felt in Aquila in the period between 1000 and 1550. The red dot identifies the event of December 03, 1315, reported in the section "Reconstruction of the Events".}\\
\\ \\
\includegraphics[width=1\linewidth]{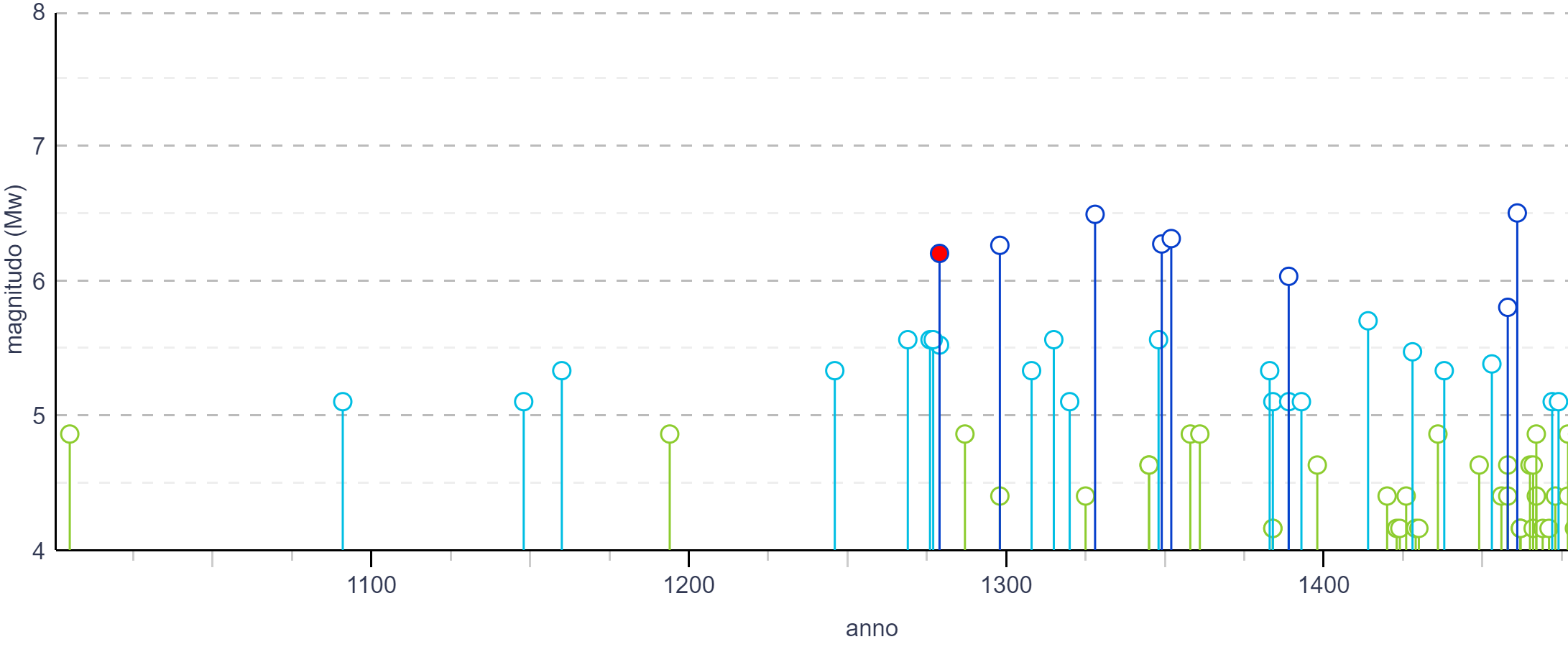}
\normalsize {Fig.21: Seismicity felt in Foligno in the period between 1000 and 1500. The red dot identifies the event of April 30, 1279, reported in the "Reconstruction of Events" section.}\\
\\ \\
\includegraphics[width=1\linewidth]{12790430_1800_000_seismicity_foligno.png}
\normalsize {Fig.22: Seismicity felt in Serravalle in the period between 1000 and 1500, with the red dot it identifies the event of April 30, 1279, reported in the section "Reconstruction of the Events.}
\\ \\
\includegraphics[width=1\linewidth]{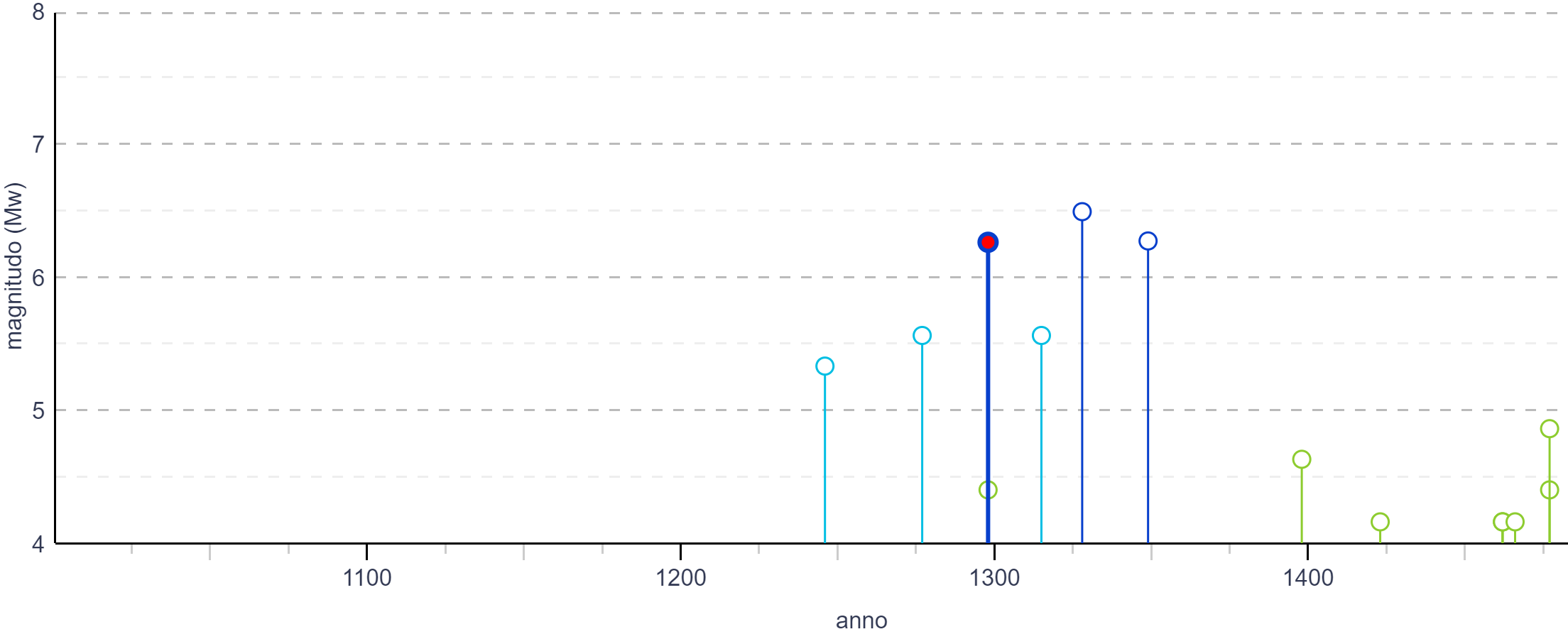}
\normalsize {Fig.23: Seismicity felt in the Rieti area in the period between 1000 and 1500, with the red dot identifying the event of December 01, 1298, reported in the section "Reconstruction of the Events.}
\\ \\
\includegraphics[width=1\linewidth]{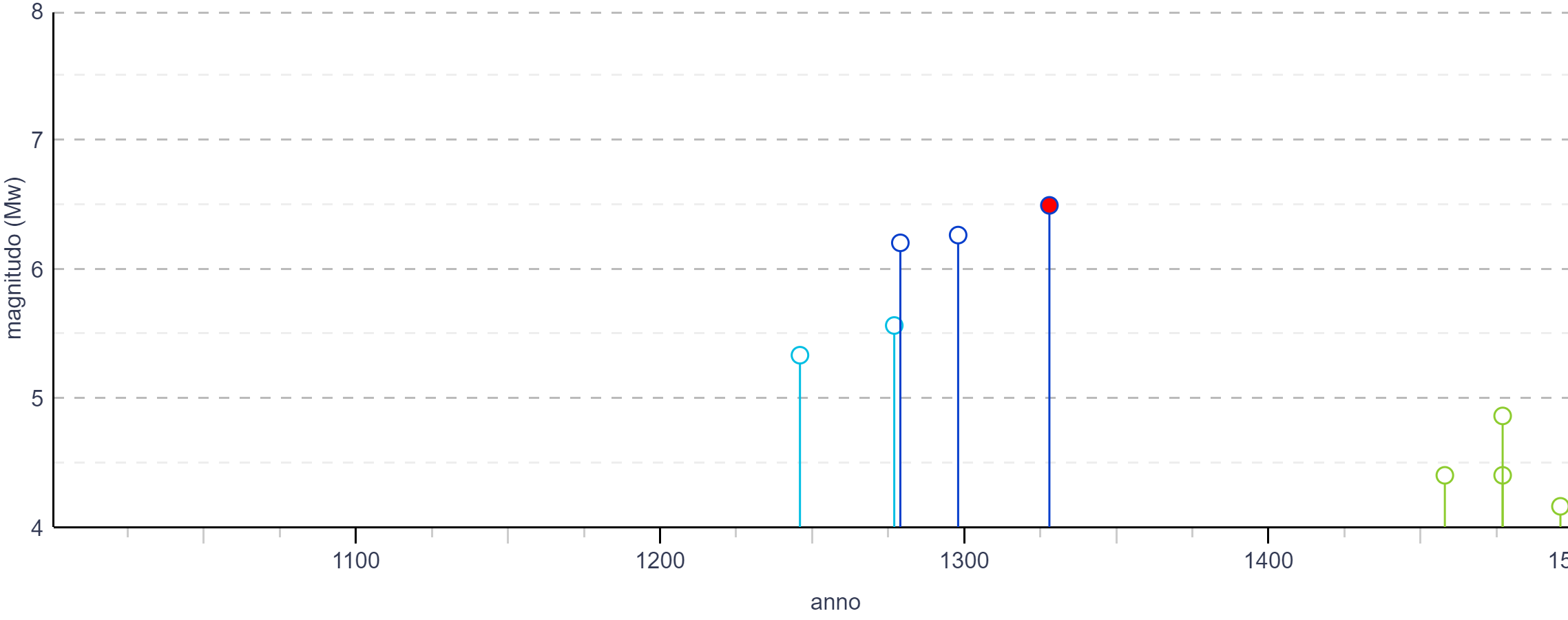}
\normalsize {Fig.24: Seismicity felt in the period between 1000 and 1500, with the red dot identifies the event of December 04, 1328, reported in the section "Reconstruction of the Events.}
\\ \\
\includegraphics[width=1\linewidth]{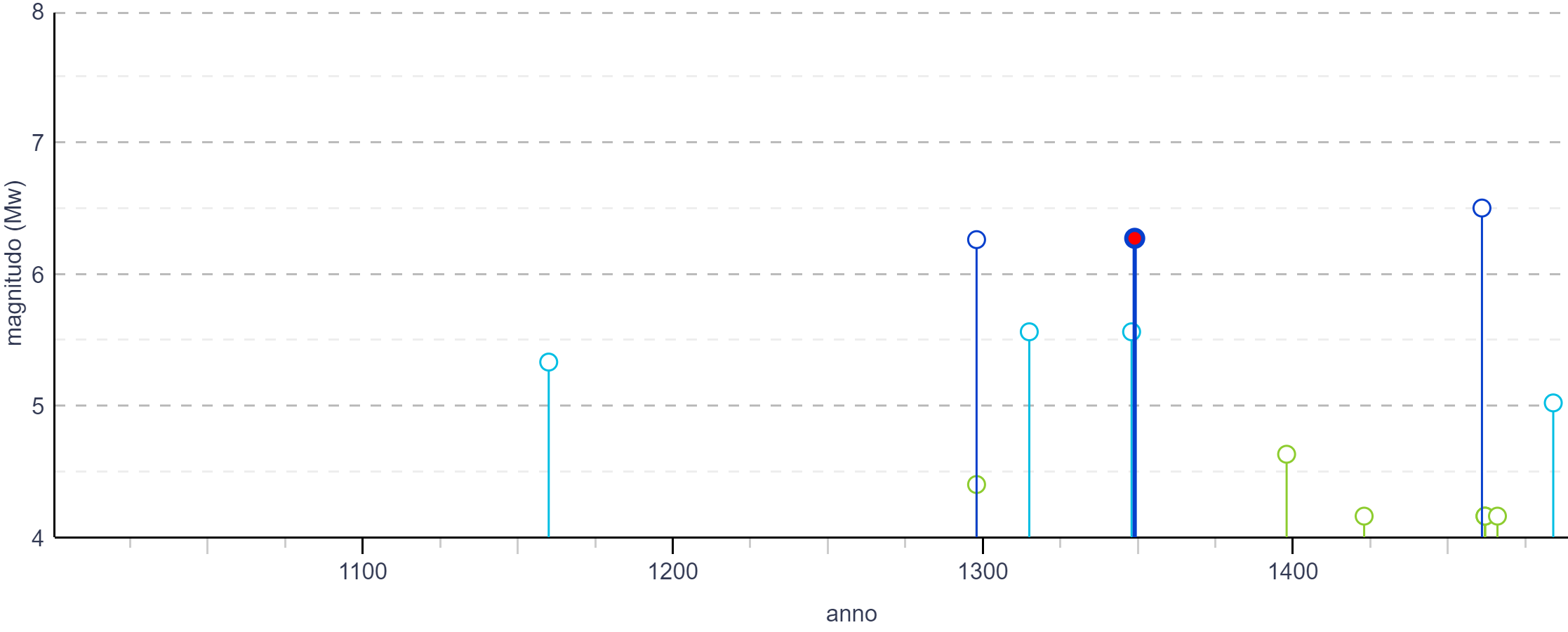}
\normalsize {Fig.25: Seismicity felt in the period between 1000 and 1500, with the red dot identifying the event of September 09, 1349, reported in the section "Reconstruction of Events.}
\\ \\
\includegraphics[width=1\linewidth]{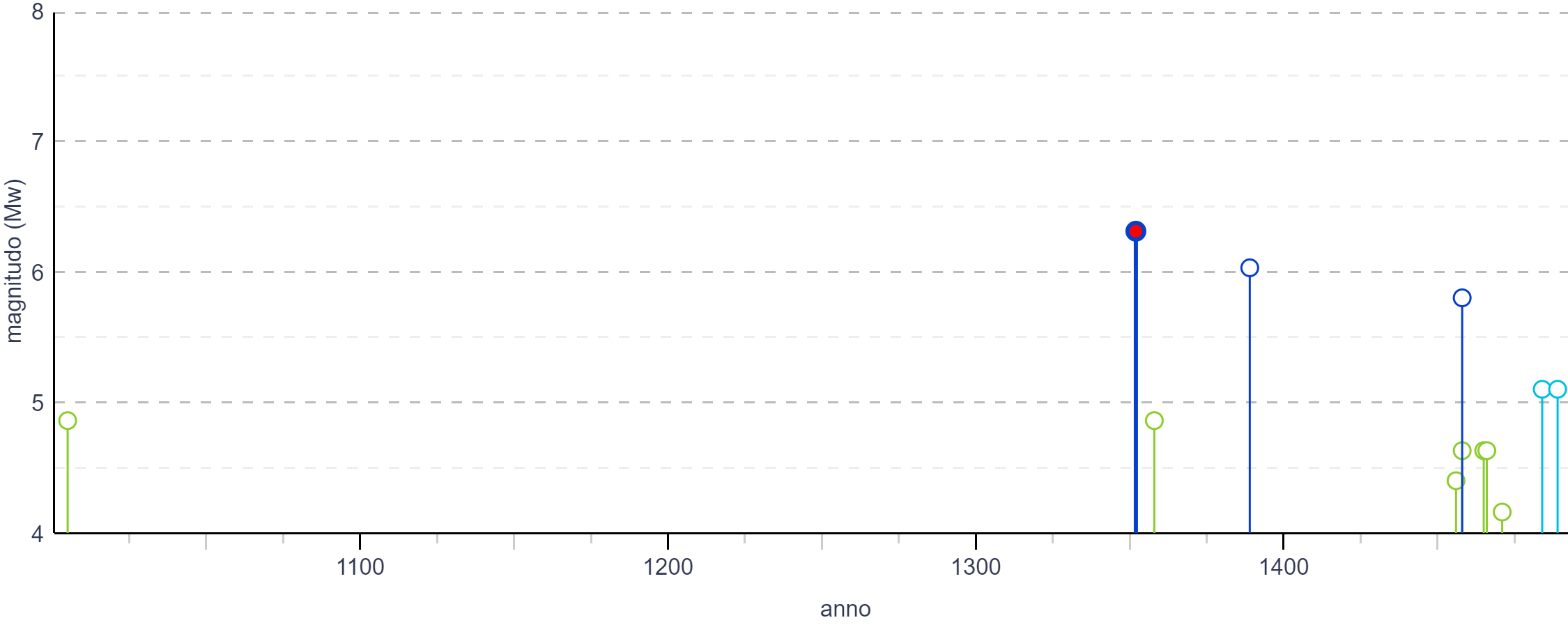}
\normalsize {Fig.26: Seismicity felt in the period between 1000 and 1500, with the red dot it identifies the event of December 25, 1352, reported in the section "Reconstruction of Events.}
\\ \\
\includegraphics[width=1\linewidth]{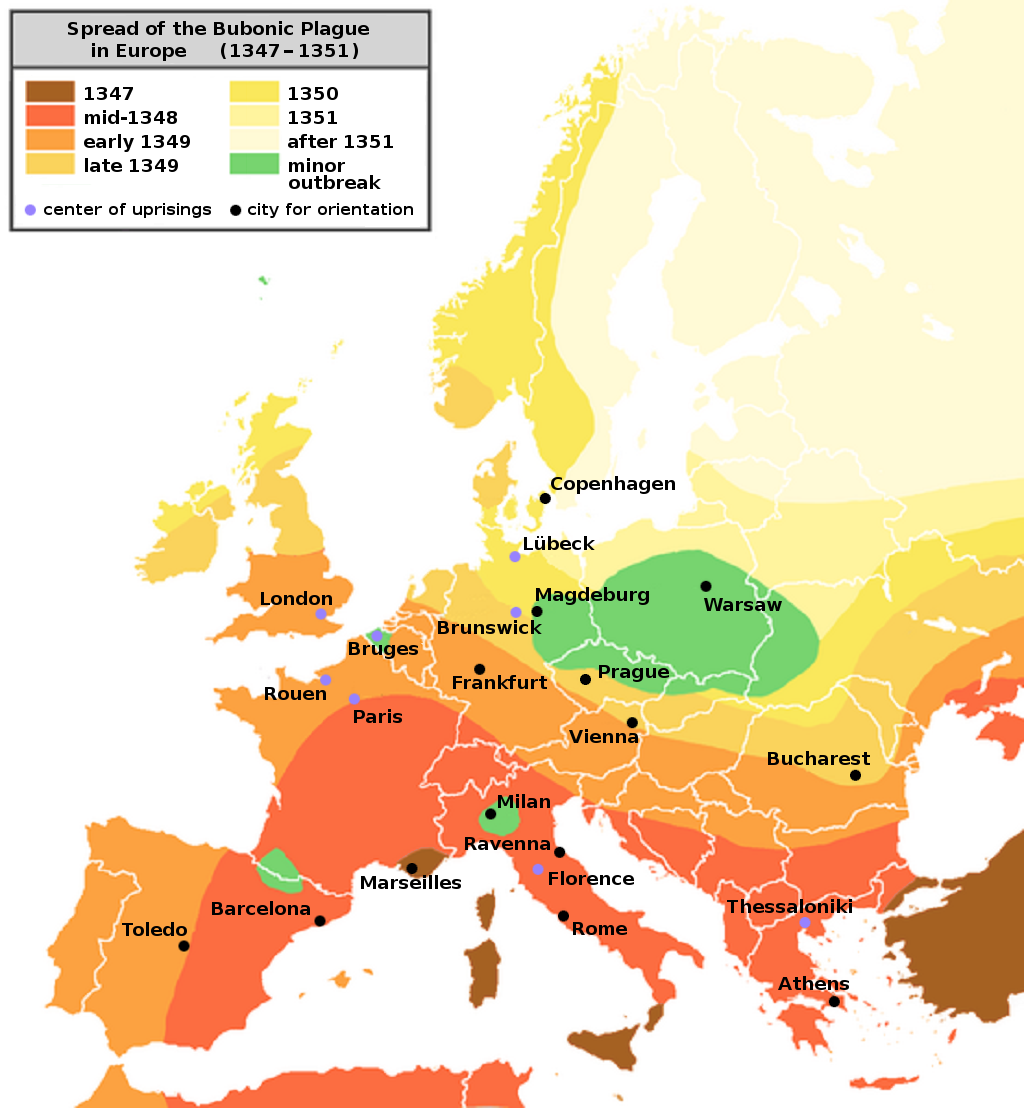} \\
\normalsize {Fig.27: Map of the spread of the plague in Europe from 1347 to 1351} {\scriptsize $^{\cite{Peste:1}}$}
\clearpage 
\centering

\end{document}